\documentclass[usenatbib]{mnras}
\usepackage[utf8]{inputenc}
\usepackage{amsmath}
\usepackage{graphicx}
\usepackage[T1]{fontenc}
\usepackage{ae,aecompl}
\usepackage{subfig}

\title[Tidally induced bars in Illustris galaxies]{Tidally induced bars in Illustris galaxies}

\author[N. Peschken and E. L. {\L}okas]{Nicolas Peschken$^{1}$\thanks{Contact e-mail:
\href{mailto:npeschken@camk.edu.pl}{npeschken@camk.edu.pl}}and Ewa L. {\L}okas$^{1}$ \\
$^{1}$Nicolaus Copernicus Astronomical Center, Polish Academy of Sciences,
Bartycka 18, 00-716 Warsaw, Poland}

\pubyear{2018}

\begin{document}
\label{firstpage}
\pagerange{\pageref{firstpage}--\pageref{lastpage}}
\maketitle

\begin{abstract}
We study barred galaxies selected from the Illustris cosmological simulation, focusing on tidally induced bars
formed from flyby interactions. To guarantee high enough resolution we focus on high mass disc galaxies
($M_*>8.3 \times 10^{10} M_{\odot}$). We find that the fraction of barred galaxies among those (21\% at redshift $z=0$)
is lower in Illustris than observed in the local Universe, and the fraction grows slightly with
redshift. The bar fraction also increases with the stellar mass and decreases with the amount of gas in the disc.
Only very few bars at redshift $z=0$ are formed in secular evolution ($\sim$ 7\%) and most of them are
triggered by external perturbers in mergers or flybys. Many of these bars disappear over time,
mostly during secular evolution, which leads to a lower fraction of bars at redshift $z=0$. We then focus on
the effect of flyby interactions on the disc and look at tidally induced bars created by a flyby, or pre-existing bars
influenced by the passage of a perturber. In the latter case, the interaction can enhance or weaken the bar. During the
interaction, the change in the bar strength occurs right after the pericentre passage. The resulting tidally induced
bars tend to be stronger than the overall bar sample in Illustris. The preferred scenario to create
or enhance a bar seems to be with a strong interaction involving a perturber on a prograde orbit. Furthermore, the
strength of the created bar grows with the strength of the interaction.
\end{abstract}


\section{Introduction}

Bars are a very common feature in late-type galaxies, their frequency in the local universe being roughly between 50\%
and 70\% (\citealt{2000AJ....119..536E}; \citealt{2009A&A...495..491A}; \citealt{2012MNRAS.423.1485S};
\citealt{2015ApJS..217...32B}; \citealt{2016A&A...596A..25L}). Besides being a very visible pattern in the stellar
structure of disc galaxies, their impact on the galactic dynamics and evolution is huge. As a non-axisymmetric
component, the bar exerts strong gravitational torques on the galactic structures and affects the angular momentum
redistribution in the whole galaxy (e.g. \citealt{2002A&A...392...83B}; \citealt{2006ApJ...645..209D}; \citealt{2007ApJ...666..189B}), driving stellar and gaseous migration in the disc
(\citealt{2010MNRAS.407.2122M, 2011A&A...527A.147M}, \citealt{2013A&A...553A.102D}; \citealt{2015A&A...578A..58H}). Understanding their formation and evolution is therefore a major goal in the study of
disc galaxies.

Bars form from disc instabilities, and numerical simulations have shown how even small perturbations can eventually
induce a bar in the centre of an axisymmetric disc (\citealt{1987MNRAS.228..635N}; \citealt{1990A&A...230...37G};
\citealt{1998ApJ...499..149M}; \citealt{2004MNRAS.347..220B}). There are thus two ways to form a bar in disc galaxies,
often referred to and opposed as \textit{nature} vs \textit{nurture}: either the bar forms in the secular evolution of
the galaxy, or it is triggered by an external perturber. While in the former case the bar slowly grows over time from
internal disc instabilities, increasing in length and slowing down (\citealt{2000ApJ...543..704D};
\citealt{2003MNRAS.341.1179A}; \citealt{2007ApJ...666..189B}; \citealt{2015PASJ...67...63O}), the latter case involves
a quick elongation of the disc, forming a so-called tidally induced bar (or tidal bar for short). This is generally due
to minor mergers (e.g. \citealt{1987MNRAS.228..635N}, \citealt{2015MNRAS.447.1774G}) or a close
passage of another galaxy (\citealt{1990A&A...230...37G}; \citealt{2014MNRAS.445.1339L}; \citealt{2014ApJ...790L..33L};
\citealt{2018arXiv180309465L}), also called a flyby, resulting in a tidal interaction creating an elongated structure
in the centre (the future bar), often with tidal extensions in the outer part. In this scenario, the bar formation does
not depend only on the characteristics of the disc, but also on the parameters of the interaction, such as the mass of
the perturber, its velocity, distance and orbit. In particular, prograde orbits have been shown to have a bigger impact
on galactic structures than retrograde orbits (\citealt{1990A&A...230...37G}; \citealt{2008ApJ...687L..13R};
\citealt{2014ApJ...790L..33L}; \citealt{2015ApJ...810..100L, 2016ApJ...826..227L}; \citealt{2018arXiv180309465L}).
However, the parameters relevant for the formation of a tidal bar are many and not easy to disentangle, so that further
work is needed to fully understand them, with the help of simulations.

Recent simulations manage to create realistic bars in disc galaxies, with different extents and morphologies (e.g. \citealt{2014MNRAS.445.1339L}; \citealt{2016ApJ...821...90A}), allowing us to study in detail the formation mechanisms of bars, their
properties and the stellar orbits constituting them. However, most of these simulations were run in isolation or, in
the case of tidal bars, in the presence of just one perturber. Although some real galaxies might be rather isolated,
most undergo constant interactions with each other, and experience many diverse encounters in their life, especially
at higher redshift (e.g. \citealt{2009A&A...507.1313H}). Multiple and repeated interactions, as well as environmental effects,
have a strong impact on the creation and subsequent evolution of a bar in the galaxy. Indeed, these constant external
perturbations can trigger the creation of a bar, or prevent it, enhance and strengthen a pre-existing bar, or on the
contrary suppress it (\citealt{2008ApJ...687L..13R}, \citealt{2018MNRAS.479.5214Z}). In this study, we aim to study bars with an emphasis
on tidal bars, their formation and evolution in a more realistic framework of a cosmological context involving diverse
interactions, using the Illustris simulation.

Illustris (\citealt{2014MNRAS.444.1518V}) is a large hydrodynamical cosmological simulation based on the moving mesh
code AREPO (\citealt{2010MNRAS.401..791S}), which is publicly available and reproduces many observational results in
the area of galaxy formation and evolution (e.g. \citealt{2015MNRAS.454.1886S}). In this study, we use the Illustris-1
simulation, which contains about 18 billion particles in a (106.5 Mpc)$^3$ volume, with a mass resolution of $4
\times 10^8 M_{\odot}$ for the dark matter and $8.1 \times 10^7 M_{\odot}$ for the gas, and follows the evolution of
the Universe from redshift 127 to 0, with 136 snapshots.

In this paper, we investigate bars in Illustris disc galaxies, in particular tidal bars formed from flyby
interactions. In section \ref{partI} we look at the general properties of bars in Illustris, while in section
\ref{partII} we focus on tidal bars triggered or affected by flyby interactions. We discuss our results in section
\ref{discuss} and conclude in section \ref{conclusion}.

\section{Barred galaxies in Illustris}
\label{partI}

\subsection{Selection of disc galaxies}
\label{discs}

The Illustris simulation provides the SubFind Subhalo catalog (\citealt{2014MNRAS.444.1518V}; \citealt{2015A&C....13...12N}) and the
SubLink merger tree, which allow to track a given galaxy (called \textit{subhalo}) over time. To look for bars in
Illustris galaxies, we first need to find disc galaxies, as we do not want to consider bars in early-type galaxies. To
do this, we used two parameters given by Illustris: the stellar circularities and the axis ratios. The circularity
parameter $\epsilon$ of a stellar particle is its angular momentum along the rotation axis, divided by the angular
momentum of the corresponding circular orbit. Therefore, disc particles are expected to have $\epsilon$ close to 1.
Illustris provides for every galaxy the fractional mass $f_{\epsilon}$ of stars with $\epsilon > 0.7$, which can
be used as a measure of the fraction of stellar mass in the disc component.
Therefore, we define disc galaxies as galaxies having more than 20\% of their stellar mass behaving kinematically as a
disc, i.e. with $f_{\epsilon} > 0.2$. We also use the axis ratios to derive the flatness of the galaxy, from the
eigenvalues of the stellar mass tensor $M_1$, $M_2$, and $M_3$, provided by Illustris. The flatness is defined as the
ratio $M_1/\sqrt{M_2 M_3}$, where $M_1$ is the lower eigenvalue. We take 0.7 as a maximum value for a galaxy to be a
disc galaxy, in addition to the $f_{\epsilon}$ constraint.

We also need to fix a lower limit for the number of particles in the galaxies we consider, in particular for the
stellar particles, to make sure the stellar disc can form a bar. After visual inspection we start with 40 000 stellar
particles as a minimum. We find that 46.7\% of galaxies with more than 40 000 stellar particles ($M_* > 3.3
\times 10^{10} M_{\odot}$) at redshift $z=0$ are disc galaxies, which makes 1232 disc galaxies.

\subsection{Bar strength and bar fraction}
\label{barstr}

We then select those 1232 disc galaxies to look for bars. To find barred galaxies, we used the $A_2$ parameter
(\citealt{2013MNRAS.429.1949A}), defined from the Fourier components:
\begin{equation}
  a_m(R)=\sum_i M_i \cos(m \phi_i)
\end{equation}
\begin{equation}
  b_m(R)=\sum_i M_i \sin(m \phi_i)
\end{equation}
\begin{equation}
  A_2(R)=\frac{\sqrt{a_2^2 + b_2^2}}{a_0}
\end{equation}
where $R$ is the cylindrical radius, $M_i$ is the mass of the $i$th stellar particle, and $\phi_i$ is its position angle.

\begin{figure*}
  \includegraphics[scale=0.3]{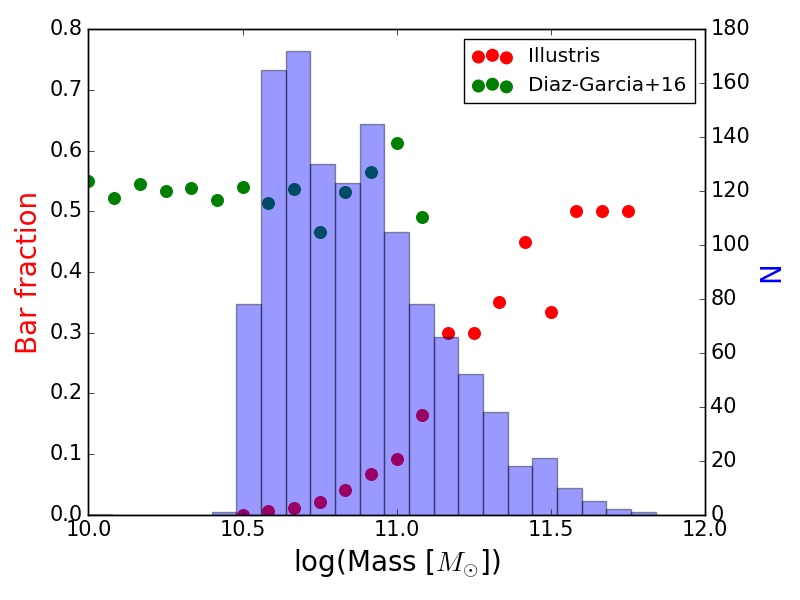}
  \includegraphics[scale=0.3]{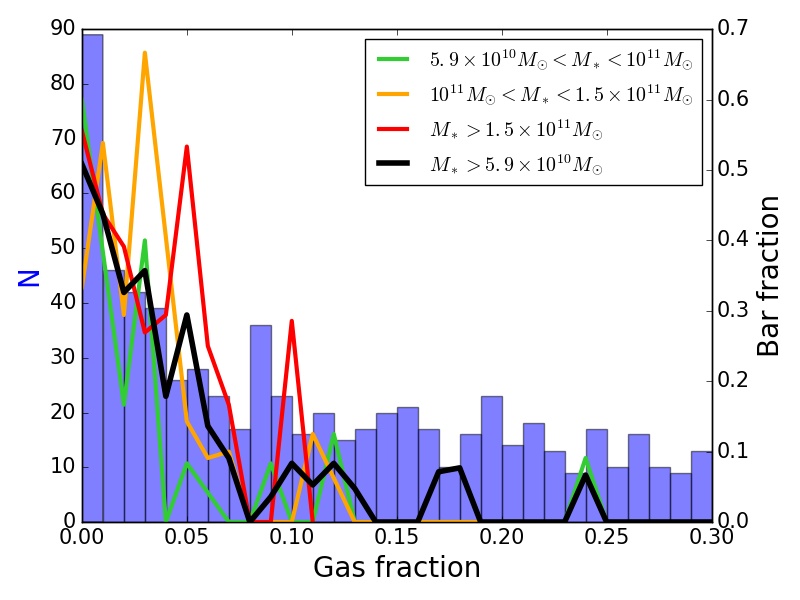}
\caption{Left panel: in red: bar fraction as a function of total stellar mass for all disc galaxies with more than 40
000 stellar particles at redshift $z=0$. In blue: histogram of the stellar masses of these galaxies. In green:
bar fraction for the 587 galaxies from \citet{2016A&A...587A.160D}. Bin size: 0.083
logarithmic units. Right panel: bar fraction as a function of the disc gas fraction at redshift $z=0$ for galaxies
in different mass bins. The green curve is for galaxies having between 70 000 and 120 000 stellar particles,
the orange one between 120 000 and 180 000 particles and the red curve above 180 000.
The black curve is for the total sample of galaxies with more than 70 000 particles. In blue: the histogram of
the distribution of corresponding gas fractions in these galaxies. Bin size: 0.01.}
\label{frac_m}
\end{figure*}

For a barred galaxy, $A_2$ as a function of radius shows a peak close to the centre corresponding to the bar
with the peak higher for stronger bars (some examples are shown later on in the subpanels of Fig.~\ref{examples}).
Therefore, we computed $A_2(R)$ for our disc
galaxies, and considered them as barred if the maximum of $A_2$ within the stellar half-mass radius is higher than
0.15. Note that a visual inspection is always needed for all the galaxies of our sample to confirm
that this peak is indeed due to a bar, and not to another non-axisymmetric component such as spiral arms or tidal
tails. We also remove from the sample those cases where it is not clear whether there is a bar or not. We find
indeed that in about 11\% of the disc galaxies the $A_2$ profile shows a peak above 0.15 that is not related to a bar
component. We define this maximum value $A=A_{2,max}$ as the bar strength to be used throughout the whole paper.

We take all disc galaxies (as defined previously) at redshift $z=0$ with more than 40 000 stellar particles (1232
galaxies) and determine the presence of bars using the value of $A=A_{2,max}$ as well as visual inspection. This allows
us to measure the fraction of barred galaxies in our sample, which we plot in Fig.~\ref{frac_m} (left panel) as a
function of the galaxy stellar mass. The total stellar mass is provided by Illustris for each galaxy, but it can
sometimes be inaccurate, as the division of stellar particles among nearby galaxies is not always reliable. We
show an example of a galaxy where the stellar particles are wrongly attributed by Illustris in
Appendix~\ref{app_mass}. We thus derived the stellar masses ourselves by fitting the stellar surface density profile
of the disc with an exponential, estimating the disc scale-length and then summing the masses of all the
stellar particles within 10 times this scale-length, and within 5 kpc vertical distance from the disc plane.

\noindent We see in
Fig.~\ref{frac_m} (left panel) that the fraction of barred galaxies increases with stellar mass, so that low-mass
galaxies ($5.9 \times 10^{10}<M_*<8.3 \times 10^{10}M_{\odot}$) almost never have bars, while more than 30\% of
high-mass galaxies ($M_*>1.5 \times 10^{11}M_{\odot}$) are barred. The bar fraction increasing with stellar
mass is consistent with many recent observational studies, such as \citet{2008ApJ...675.1141S},
\citet{2012MNRAS.423.1485S}, \citet{2014MNRAS.438.2882M}, \citet{2016A&A...587A.160D}, \citet{2015A&A...580A.116G},
\citet{2016A&A...595A..67C}, \citet{2017ApJ...835...80C}. In Fig.~\ref{frac_m} we also plotted the results of
\citet{2016A&A...587A.160D} for comparison with observations, although it is very difficult to find data for high mass
galaxies as in our sample. We see that at stellar masses between $10^{10.5}$ and $10^{11} M_{\odot}$ we find much less
bars than expected from observations.

Since at low masses we do not find bars in Illustris galaxies, we will from now on restrict our analysis to
galaxies with stellar masses higher than $8.3 \times 10^{10}$ $M_{\odot}$ to be able to study the bar frequency, and
will refer to this sample as our \textit{default sample} (509 galaxies). This mass threshold corresponds to 100
000 stellar particles.

Furthermore, we look for a dependence of the bar fraction on the gas content in the disc. We compute the gaseous mass
in the disc in a similar way as was done for the stellar mass and derive the gas fraction in the disc, for our sample of disc galaxies. We plot in
Fig.~\ref{frac_m} (right panel) the bar fraction as a function of the gas fraction in different mass bins and find a clear decreasing trend: the more
gas the disc has, the less likely it is to host a bar, for the total sample and for all mass bins. Note that the bar fraction decreases clearly up to $\sim$ 8\% of gas, and is then flat and very close to 0. The increasing trend of the bar fraction with decreasing gas
content is also consistent with observations (\citealt{2012MNRAS.424.2180M}; \citealt{2013ApJ...779..162C};
\citealt{2017ApJ...835...80C}), and with the trend of gas weakening or preventing the formation of bars seen in
simulations (\citealt{1993A&A...268...65F}; \citealt{2010ApJ...719.1470V}; \citealt{2013MNRAS.429.1949A}).

\subsection{Redshift dependence}

\begin{figure}
\includegraphics[scale=0.3]{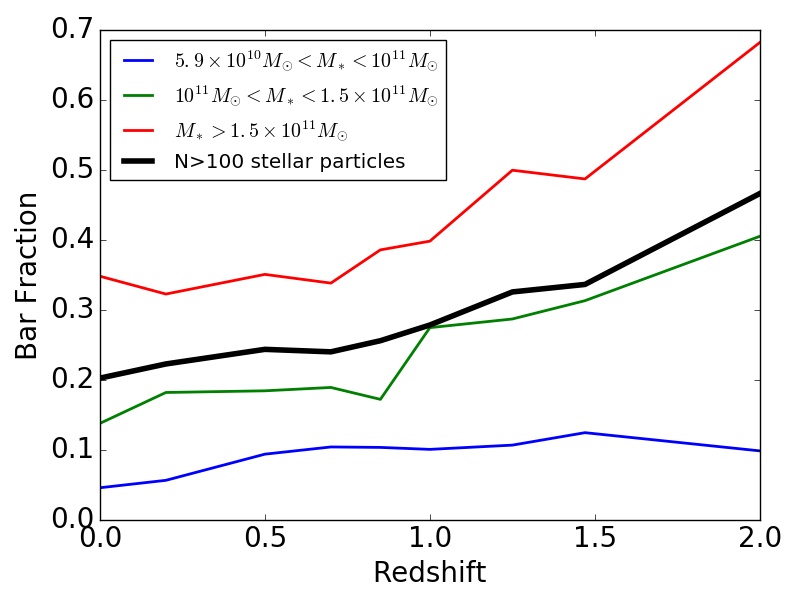}
\caption{Bar fraction as a function of redshift for different ranges of stellar mass. The black curve is for all
disc galaxies with more than 100 000 particles (default sample). The blue curve is for galaxies having between 70 000 and 120 000 stellar particles, the green one between 120 000 and 180 000 particles and the red curve above 180 000.}
\label{frac_z}
\end{figure}

We repeated the analysis of the previous subsection at different redshifts (characteristics of the
samples at different redshifts can be found in Table~\ref{table}) and found similar results: increasing trend of the
bar fraction with stellar mass, and decreasing trend with gas fraction. This allowed us to look at the bar fraction as
a function of redshift shown in Fig.~\ref{frac_z}. We find an increasing trend with redshift, which is surprising, as
most studies find less bars at high redshift (\citealt{2008ApJ...675.1141S}; \citealt{2014MNRAS.438.2882M};
\citealt{2014MNRAS.445.3466S}). Note that a similar trend is visible in different mass bins for our sample, which means
that it is not related to the resolution, except perhaps for the lowest-mass galaxies.

However, our sample
consists of massive galaxies ($M_*>8.3 \times 10^{10}M_{\odot}$), while the decrease of the bar fraction with redshift
is mostly observed for low mass galaxies (e.g. \citealt{2008ApJ...675.1141S}; \citealt{2012ApJ...757...60K}). For more
massive galaxies, the bar fraction has been found to be more or less constant with redshift
(\citealt{2008ApJ...675.1141S}; \citealt{2004ApJ...615L.105J}; \citealt{2007ApJ...659.1176M}), or even increasing. In
particular, \citet{2010MNRAS.409..346C} found that for high mass galaxies ($M_* > 10^{11}M_{\odot}$)
the fraction of bars
tends to increase with redshift, up to $z \sim 0.6$, with values similar to our results. For the galaxies with at
least 100 000 stellar particles (our default sample, $M_*>8.3 \times 10^{10}$ $M_{\odot}$), we see that we have about
21\% of galaxies hosting bars at $z=0$. Looking at the disc gas fraction in galaxies, we find an increase with
redshift, as expected.

\begin{table}
  \caption{Description of the different samples used in the paper, for galaxies with more than 100 000 stellar particles.}
  \begin{tabular}{|c|c|}
    \hline
    Sample & Number of galaxies  \\
    \hline \\
    Default sample: disc galaxies at $z=0$  & 509\\
    Barred discs at $z=0$ & 108 (21.2\%) \\
    Disc galaxies at $z=0.5$ &  505\\
    Barred disc at $z=0.5$ & 123 (24.4\%)\\
    Disc galaxies at $z=1$ & 352\\
    Barred disc at $z=1$ & 98 (27.8\%)\\
    Disc galaxies at $z=2$ & 105 \\
    Barred disc at $z=2$ & 49 (46.7\%)\\
    Interaction sample (flybys \& fly-throughs)& 121  \\
    Flyby sample (only flybys)& 79 \\
    Bar creations (interactions creating a bar)& 50 \\
    Reference sample (no interaction) & 70 \\
    \hline
    \label{table}
  \end{tabular}
  \end{table}

\begin{figure}
\includegraphics[scale=0.3]{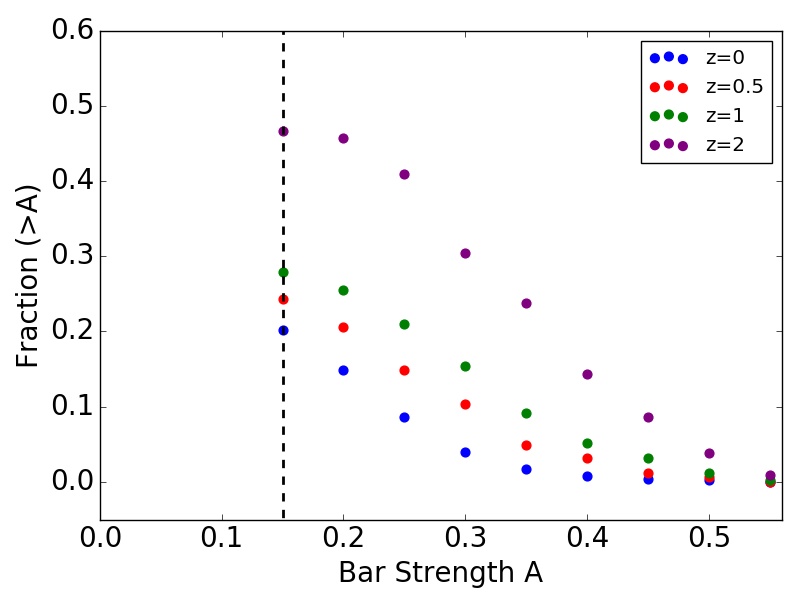}
\caption{Cumulative fraction of barred galaxies among disc galaxies with more than 100 000 stellar particles (default sample), as a
function of their bar strength, for different redshifts. The dashed vertical line corresponds to 0.15, which we fixed
as the minimum for a galaxy to be considered as barred. The points on this line thus give the total fraction of barred
galaxies.}
\label{frac_S}
\end{figure}

Measuring the strength of the bars in the galaxies of our default sample, we find the strength average of
$A=0.24 \pm 0.07$ at redshift $z=0$.
The distribution of the corresponding bar strengths can be found later on in Fig.~\ref{hists} (top left panel),
with a comparison to observations using \citet{2016A&A...587A.160D}, who also compute the bar strength from
Fourier decomposition, so the values are directly comparable. We find that our bar strengths, although
consistent with observations, tend to be in the weaker end of values observed in the local Universe. In
Fig.~\ref{frac_S} we plot the cumulative fraction of galaxies having a given strength. We see that most bars at low
redshift are relatively weak, with a bar strength below 0.3. We found that the mean bar strength increases with
redshift, which seems mainly due to the fact that at higher redshifts the bars are less symmetric and more elongated
and perturbed. The increase of the bar fraction with redshift is striking here, and seems unrelated to whether we
choose to consider weaker or stronger bars.

We tried to take into account more early-type galaxies that we did not classify as disc galaxies in
our sample to see if this increasing trend with redshift was a selection effect. We thus constructed a sample of all
the galaxies with more than 100 000 particles that were not selected in our disc galaxies sample, i.e. with either
$f_{\epsilon}<0.2$ or a flatness higher than 0.7. We then looked at the bar fraction in this sample of galaxies as in
section \ref{barstr}. We still found an increasing trend of the bar fraction with redshift in this sample. Therefore,
the increasing trend of the bar fraction with redshift is not due to a selection effect.

\begin{figure*}
  \includegraphics[scale=0.16]{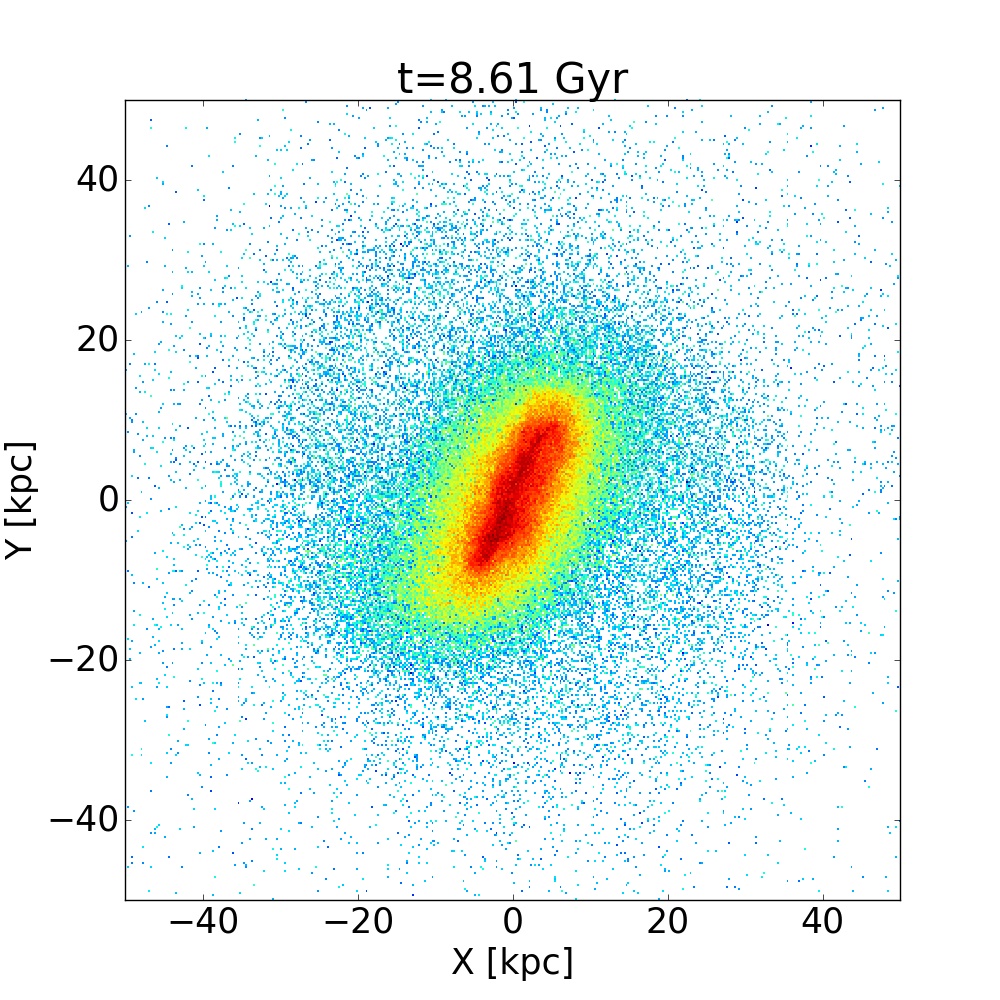}
  \includegraphics[scale=0.16]{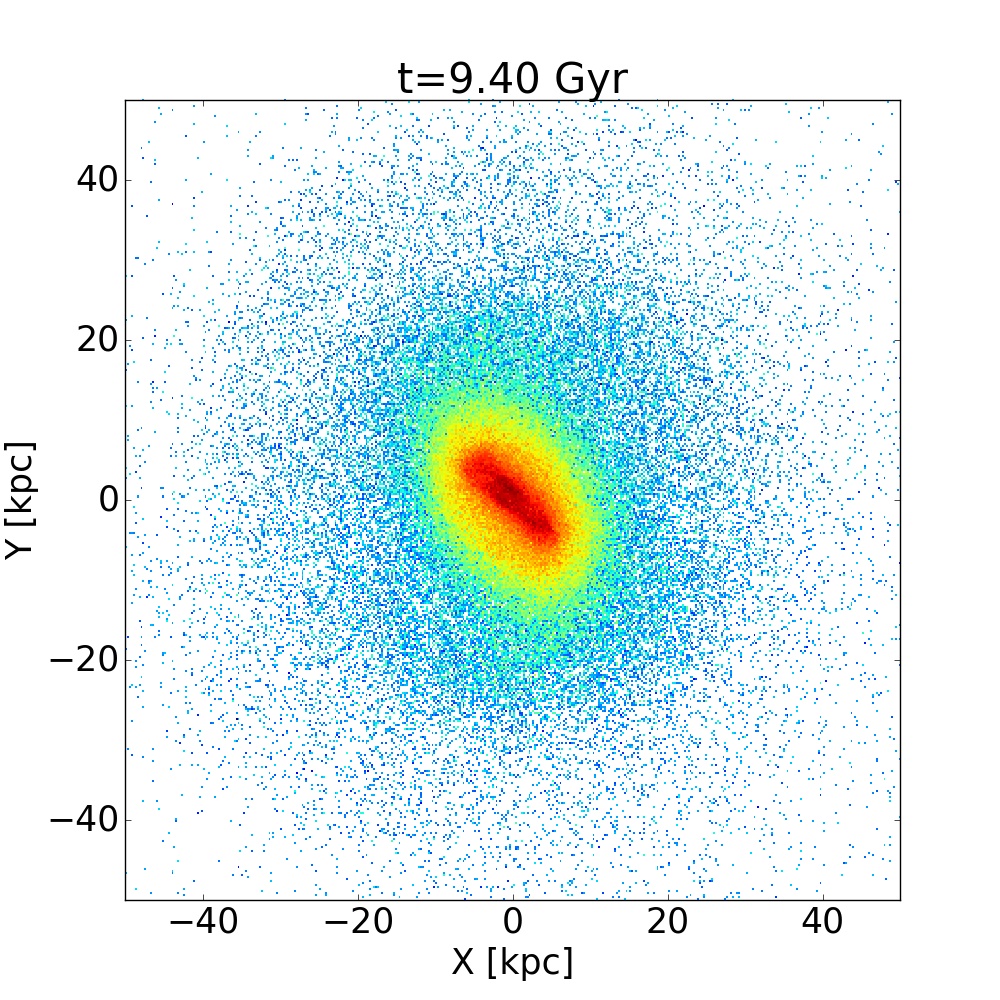}
  \includegraphics[scale=0.16]{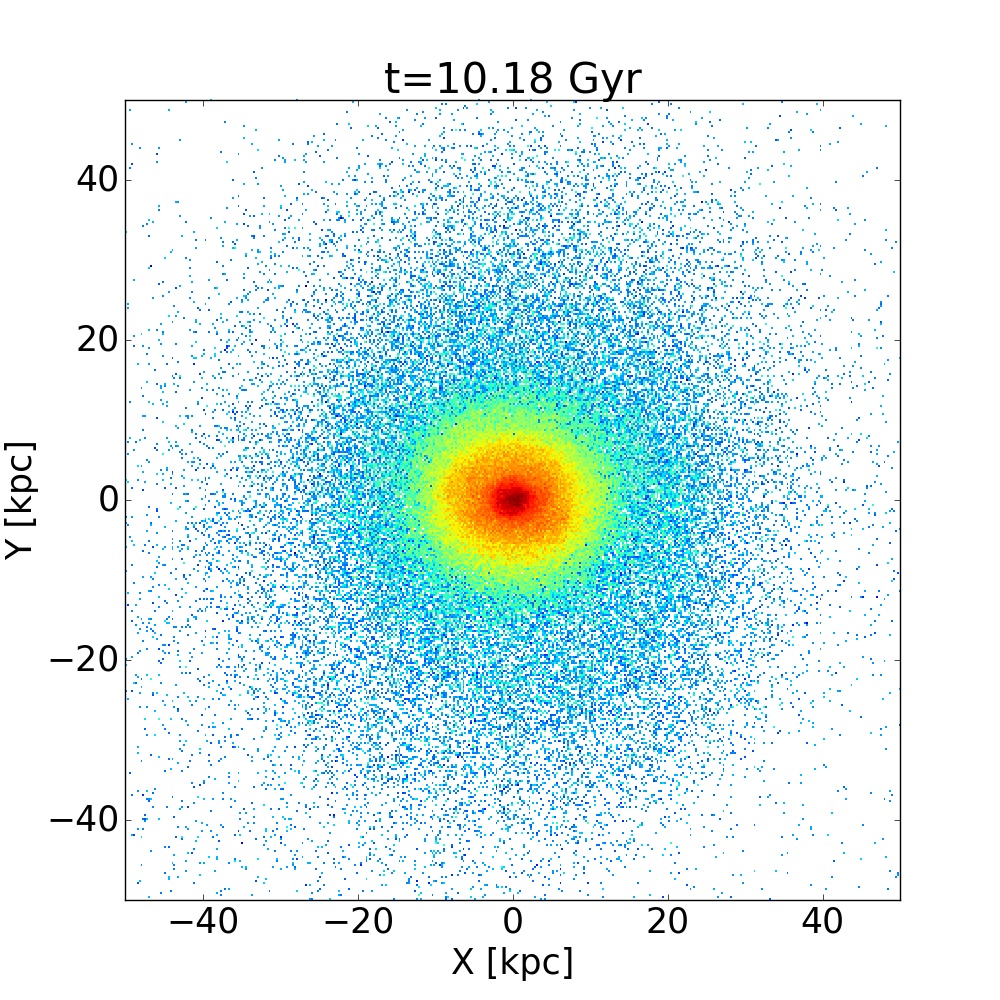}
\caption{Disappearance of a bar in secular evolution of a disc galaxy viewed face-on. The three panels show
three stages of evolution in time.}
\label{ex_disp}
\end{figure*}

We now try to understand the increasing bar fraction with redshift, by following bars over time. Indeed,
if the bar fraction decreases over time, it means that bars must disappear in one way or another. We take all the barred
disc galaxies at redshift $z=0.5$ (123 galaxies), and follow their evolution over time, to see what happens to the bar.
A significant number of galaxies (55) cannot be tracked as they fall out of the category of disc galaxies by redshift
$z=0$, probably due to minor or major mergers or heating of the disc. However, we have 68 barred galaxies at redshift
$z=0.5$ that still are disc galaxies at redshift $z=0$. Among those galaxies, 25 bars disappear during the evolution.
By tracking back the mechanism of the bar disappearance and removing three unclear cases, we find that 4 bars are short
lived, i.e. they disappear within a few snapshots after their formation. Furthermore, 4 cases correspond to mergers
(both minor and major) or interactions, and 14 to a secular decrease of the bar strength until complete disappearance
of the bar. We have checked in the latter case if the disappearance could be due to the passage of dark
satellites, but it does not appear to be the case. Although there are numerous dark haloes everywhere in Illustris,
their total mass is very low ($\sim 10^8 - 10^9 M_{\odot}$), and a hypothetical impact of the collision
would affect the bar strength abruptly, instead of the slow secular decrease we observe over time.

It thus seems that most bars are dissolved in the secular evolution of the galaxy, without external
perturbation; an example is shown in Fig.~\ref{ex_disp}. One sees a very strong bar shrinking by itself, until complete
disappearance. This is surprising, as bars are expected to grow stronger over time in the secular evolution phase (e.g.
\citealt{2013MNRAS.429.1949A}). We repeated this analysis with the barred galaxies at redshift $z=1$, and found similar
results, although the fraction of mergers destroying the bar is higher, as expected since there are more mergers at
high redshift. Furthermore, even for bars surviving over time between redshift $z=0.5$ and $z=0$, an important
fraction of them tends to decrease in strength over time, without external perturbation.

Therefore, we conclude
that the bar fraction decreasing over time in Illustris is mostly due to the bars being suppressed in the secular
evolution phase of the galaxies, with the bar strength decreasing over time. This also explains why our low redshift
bars are weaker than the high redshift ones, and than in observations (see Fig.~\ref{hists}, top left panel).

\subsection{Bar length}

The $A_2$ profile can also be used to estimate the bar length, although this is not as straightforward. The end of
the bar is expected to correspond to a drop in $A_2$, so the bar length could be defined as the radius for which $A_2$
goes below some threshold value. This threshold could be defined as an absolute value, or as a fraction of the bar
strength $A=A_{2,max}$ (e.g. \citealt{2002MNRAS.330...35A}; \citealt{2017MNRAS.469.1054A}). However, in some cases there is no strong drop of
$A_2$ at the bar end, because other features at the end of the bar keep $A_2$ at a relatively high level. This is in
particular the case when tidal tails appear in the galaxy due to external perturbations such as galaxy interactions
we will investigate in section~\ref{partII}. Therefore, we define the bar length as the radius at which the peak of
$A_2$ occurs. Although this does not correspond to the real length of the bar but rather to its lower limit, it provides us with a typical radius of
the bar, is easily reproducible and consistent for all different types of barred galaxies in our sample. Making a comparison with a visual estimation of the extent of the bar, we found that our values correspond to roughly half of the real bar extent.

We find a mean bar length of 4.7 $\pm$ 2.0 kpc at redshift $z=0$ in our default sample of disc galaxies. The
corresponding distribution of the lengths can be found later on in Fig.~\ref{hists} (bottom left panel). We find no
clear trend of the bar length with the stellar mass or with the gas fraction in the disc. However, it seems that the
bars tend to be longer at higher redshifts.
Again, this seems to be due
to the fact that at higher redshifts our discs appear to be less stable and exhibit unstable bars, longer and more
perturbed.

\subsection{Bar pattern speed}
\label{patt}

We also derived the bar pattern speed of our barred galaxies. Since the snapshot frequency in Illustris is too low
(around 150 Myr between two subsequent snapshots) to be able to simply derive the pattern speed from the evolution
of the bar position angle over time, we used the Tremaine-Weinberg method (\citealt{1984ApJ...282L...5T}). This method
uses the surface brightness and the radial velocities of the galaxy, and works both in simulations and observations.
The bar pattern speed $\Omega_p$ is then given by:
\begin{equation}
\Omega_p \sin i = \frac{\int_{-\infty}^{+\infty}h(Y) {\rm d} Y \int_{-\infty}^{+\infty} \Sigma(X,Y) V {\rm d} X}
      {\int_{-\infty}^{+\infty}h(Y) {\rm d} Y \int_{-\infty}^{+\infty} \Sigma(X,Y)X {\rm d} X}
      \label{eqTW}
\end{equation}
where $i$ is the inclination angle of the galaxy, $h$ is a weight function, $\Sigma$ is the stellar density in the disc
plane, $V$ are the stellar velocities, and $X$, $Y$ are the coordinates in the plane of the sky. We take a two slit
offset function as the weight function, as recommended by \citet{1984ApJ...282L...5T}: $h(Y)=1/(1+((Y-Y_0)/n_w)^2 -
1/(1+((Y+Y_0)/n_w)^2$, where $n_w$ is a weighing integer, fixed here at 20, and we take the height of the slice
to be equal to the bar length. We use an inclination angle of the disc plane of $i=45$ degrees.

We performed a test of this method on an $N$-body simulation of an isolated Milky Way-like galaxy from
\citet{2016ApJ...826..227L}, as described in Appendix~\ref{app_TW}. The histogram of the pattern speeds
in our default sample can be found later on in Fig.~\ref{hists} (top right panel), with a comparison to
observations from \citet{2011MSAIS..18...23C} who also uses Tremaine-Weinberg, and \citet{2017ApJ...835..279F} who use
the Font-Beckman method (\citealt{2011ApJ...741L..14F}). We find reasonable values for the pattern speeds at redshift
$z=0$ (mean value $14.2  \pm 6.4$ km s$^{-1}$ kpc$^{-1}$), although in the lower end of the values observed in the
local Universe.

\subsection{Origin of bars}
\label{origin}

We now try to answer the question how the bars present in Illustris at redshift $z=0$ were formed. We take the barred
galaxies of our default sample at this redshift (108 galaxies), and follow them backward in time to identify the
mechanism responsible for the creation of these bars. In particular, we look for cases where the bar was formed by a
merger or a flyby interaction. To do this, we compute the bar strength as a function of time and look for the moment
where the bar strength drops below 0.15 (still going backwards). We also inspect the galaxies visually to determine
when the bar is actually created. Once the formation moment is found, we check whether there is a clear galaxy
interaction involved in it (as checked visually for galaxies with stellar masses higher than
$8 \times 10^8 M_{\odot}$ in a cube of (500 kpc)$^3$ centered on the galaxy) or if it
was formed in secular evolution. We remove from our sample 26 cases where the bar formation mechanism is not easy to
point out, e.g. when there are several interactions occurring at the same time, or if it is not clear whether the bar
was triggered by an external galaxy or not. We find that 50\% of bars resulted from a minor or major merger
(in the latter case the bar forms after the major merger is complete and the disc is created)
 or multiple mergers, 42.7\% from a flyby, while the rest appear to be formed in secular evolution
(7.3\%). Therefore, interaction events seem to be the most effective way to form a bar, and we find very few cases of
bars formed in isolation in Illustris.

Furthermore, we examine the properties of the disc in these isolated galaxies before the bar forms. Bars formed in
secular evolution are expected to arise more easily in discs dominated by the stellar component than in dark matter
dominated ones, as the dark matter halo stabilizes the disc against bar formation (\citealt{1976AJ.....81...30H};
\citealt{1982MNRAS.199.1069E}).
As a rough measure of the disc's susceptibility to bar formation we use the disc fraction, which we define as the ratio
of stellar mass to the dark matter mass inside one disc scalelength. We find that all the cases where the bar appears
during the secular evolution of the galaxy have values of the disc fraction higher than two, i.e. there is at least
twice as much stellar mass than dark matter in the central region of these galaxies. This is thus consistent with the
disc being prone to form a bar, although a bigger sample of bars formed in isolation would be needed to confirm this
trend.

The fact that there are very few bars formed in secular evolution might be due to the rather high gravitational
softening length in Illustris ($0.7 h^{-1}$ kpc) which can prevent disc instabilities because of poor resolution of
small scale systems (\citealt{2007MNRAS.375...53K}). This issue has been discussed in the context of galaxy formation
in Illustris by \cite{2015MNRAS.447.2753T}.

Therefore, it seems than in Illustris, bars are mostly formed at higher redshift by galaxy interactions, but
are then suppressed over time. This thus explains why at high redshift, when interactions are numerous, one finds more
bars than at low redshift, when many bars have disappeared in the secular evolution of the galaxies.

\section{Flyby driven bars}
\label{partII}

\subsection{Selection of the sample}
\label{sample}

\begin{figure*}
    \includegraphics[scale=0.16]{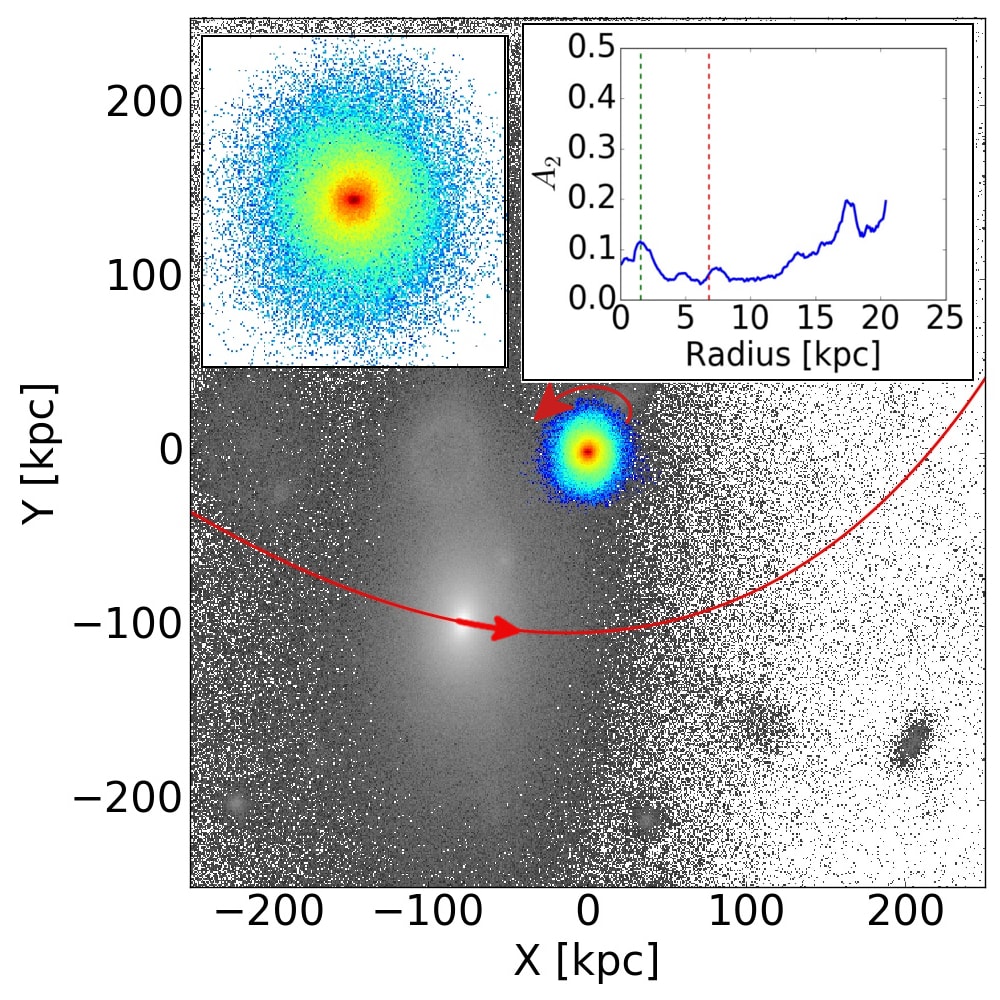}
    \includegraphics[scale=0.16]{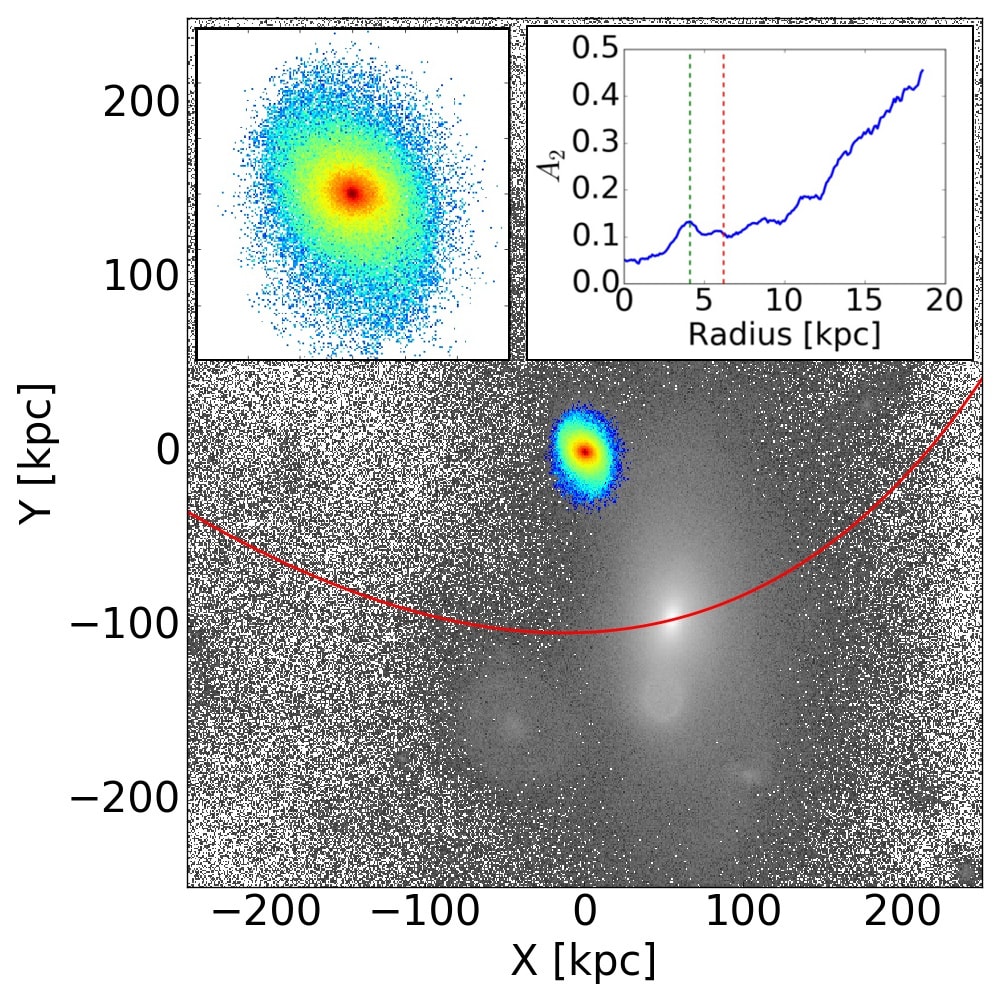}
    \includegraphics[scale=0.16]{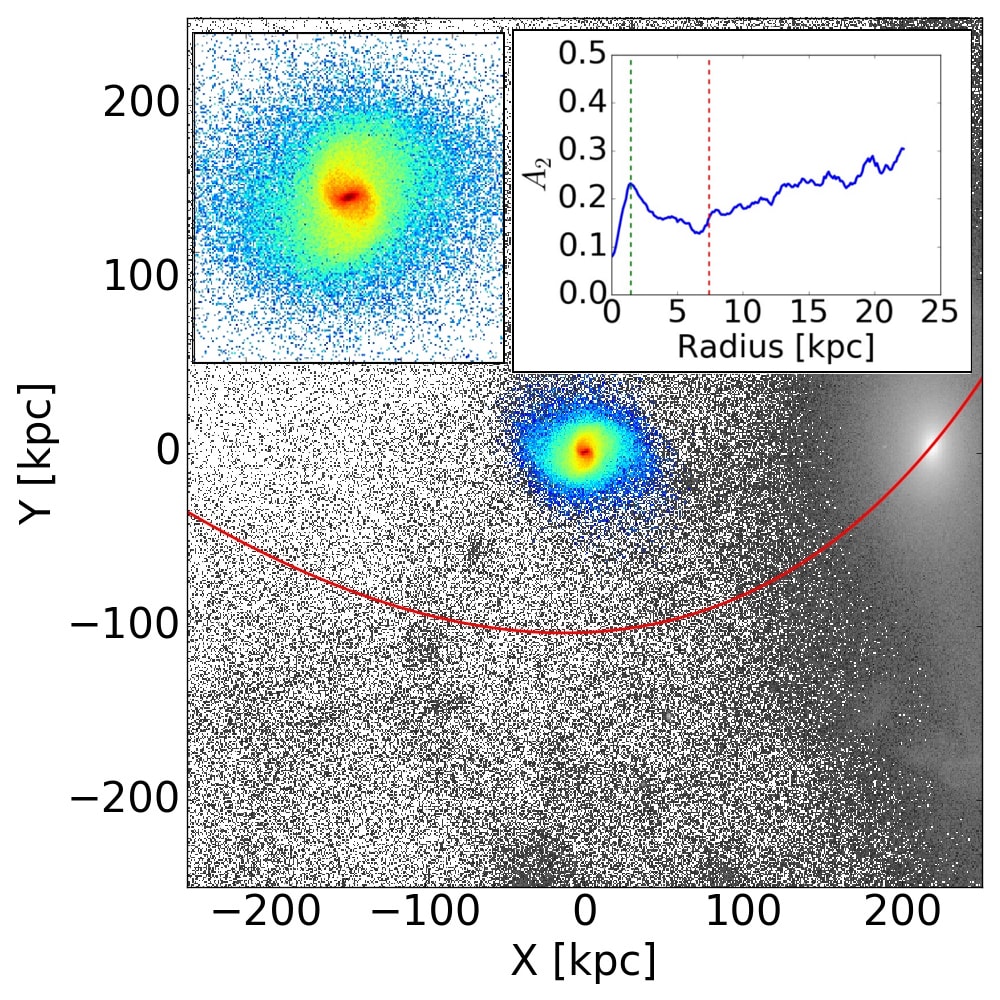}

    \includegraphics[scale=0.16]{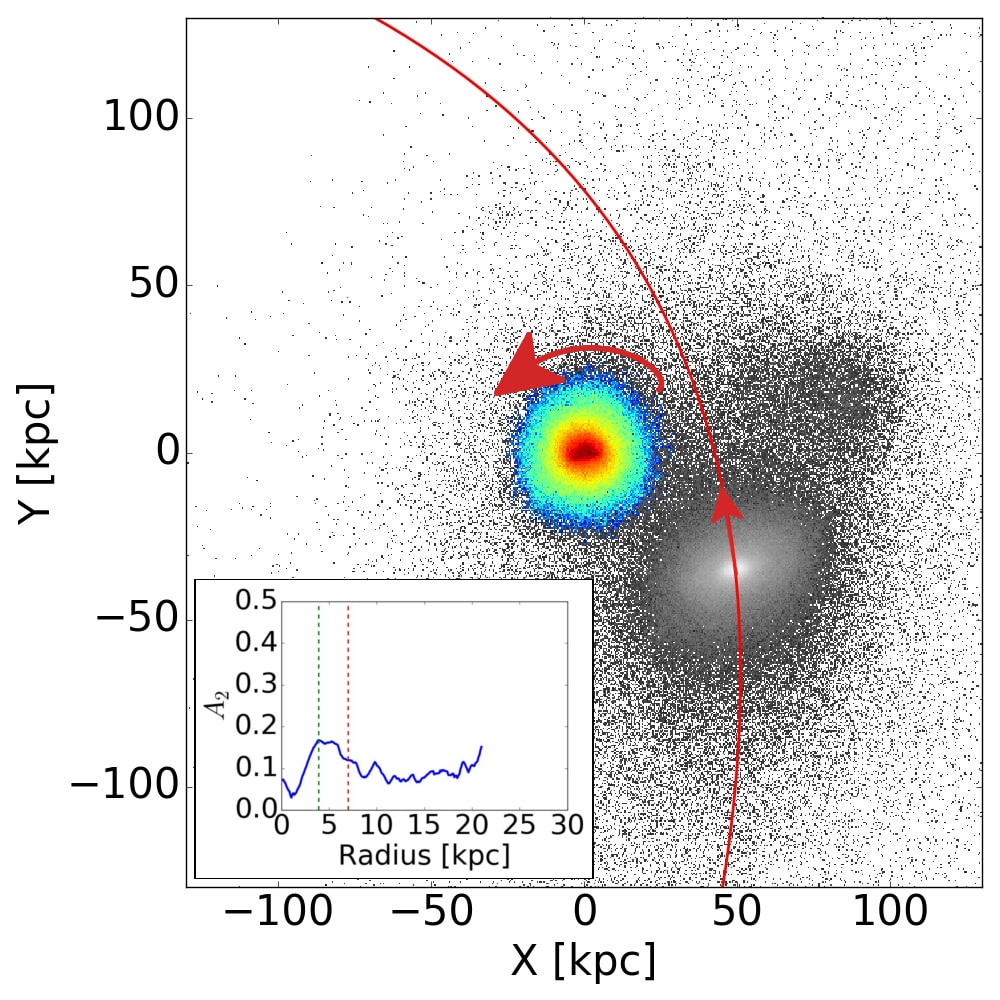}
    \includegraphics[scale=0.16]{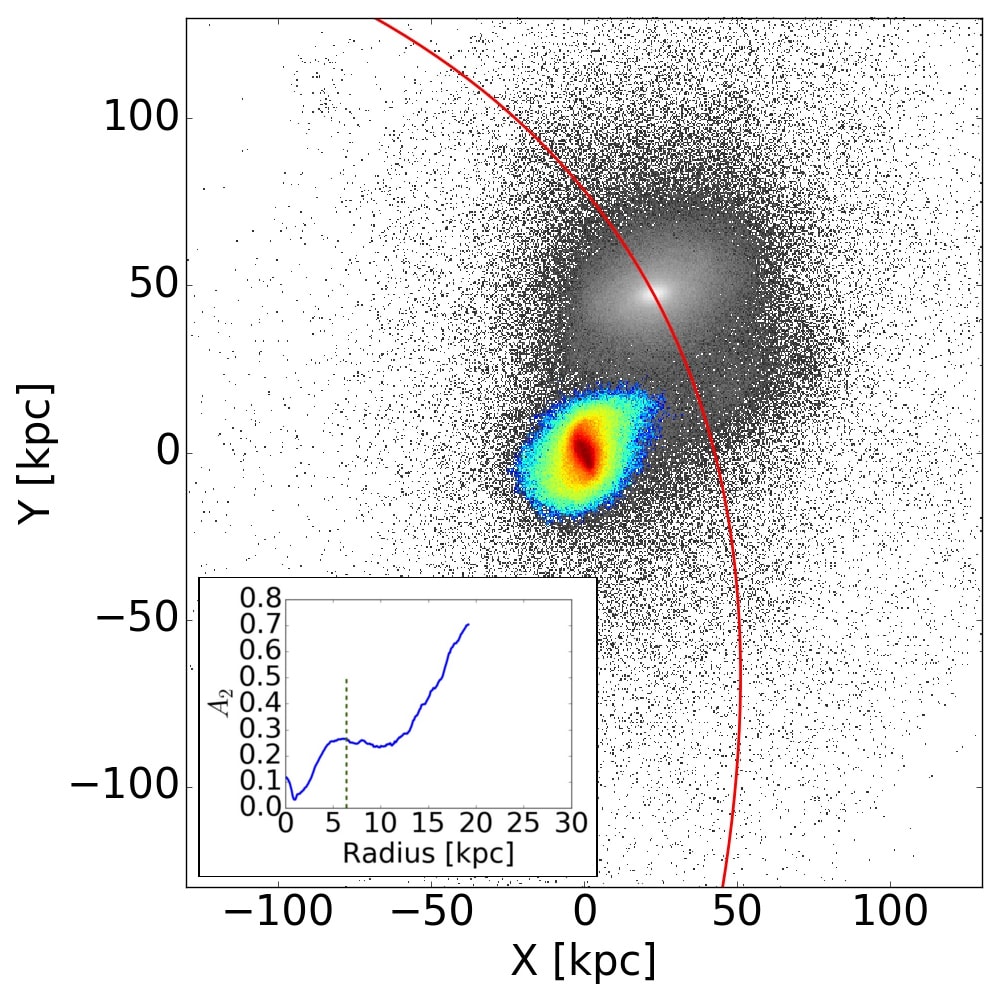}
    \includegraphics[scale=0.16]{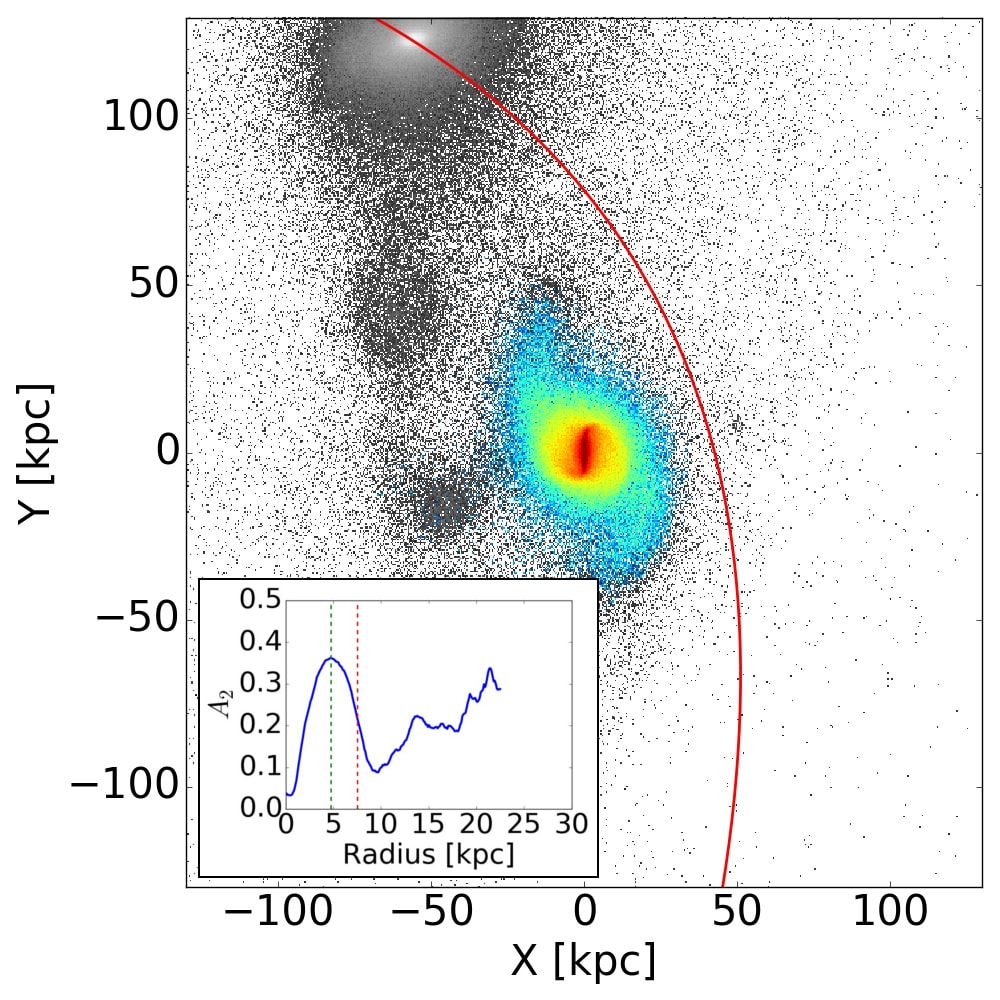}

    \includegraphics[scale=0.16]{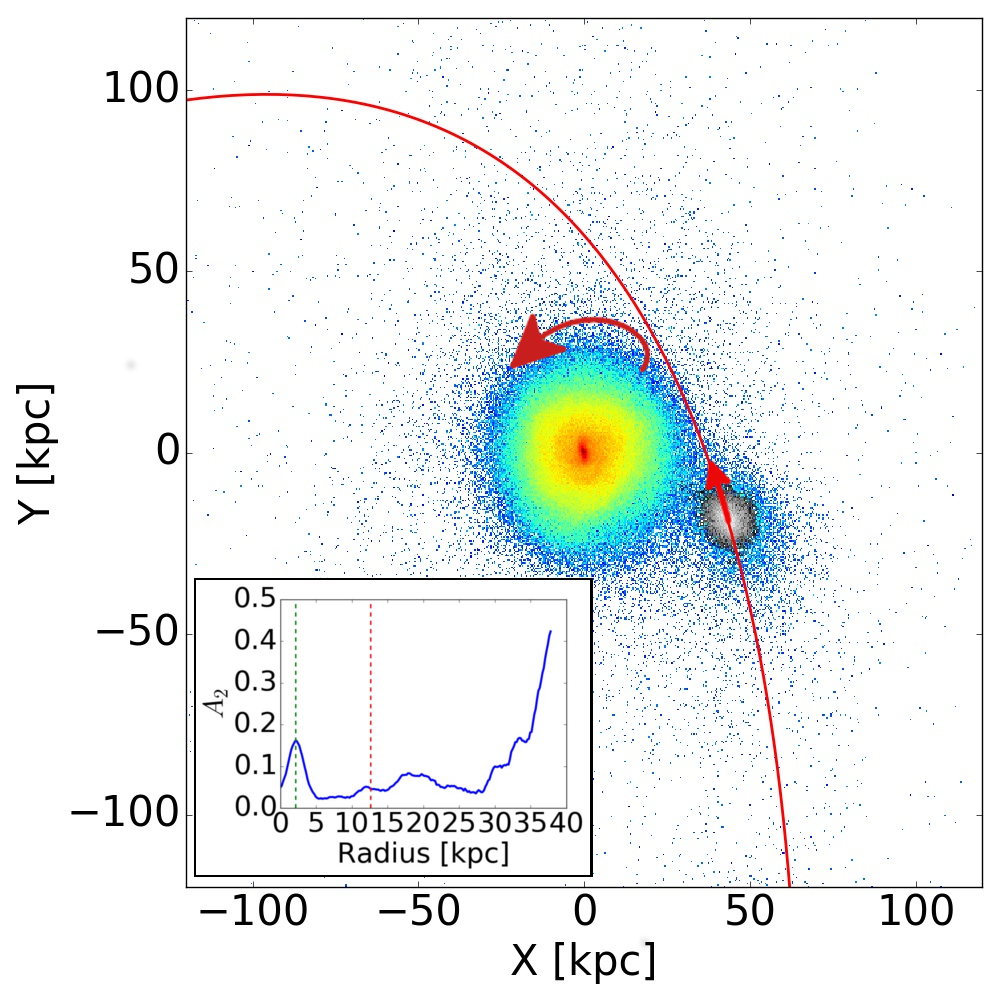}
    \includegraphics[scale=0.16]{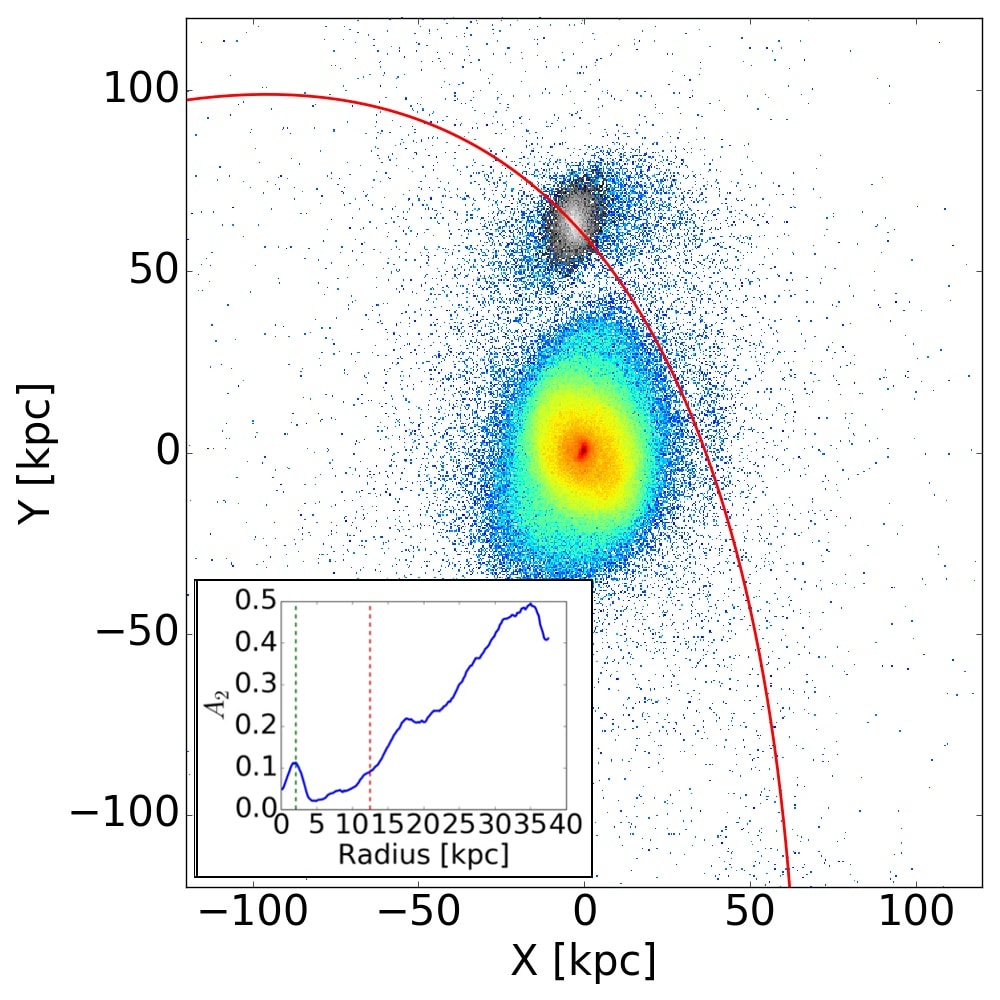}
    \includegraphics[scale=0.16]{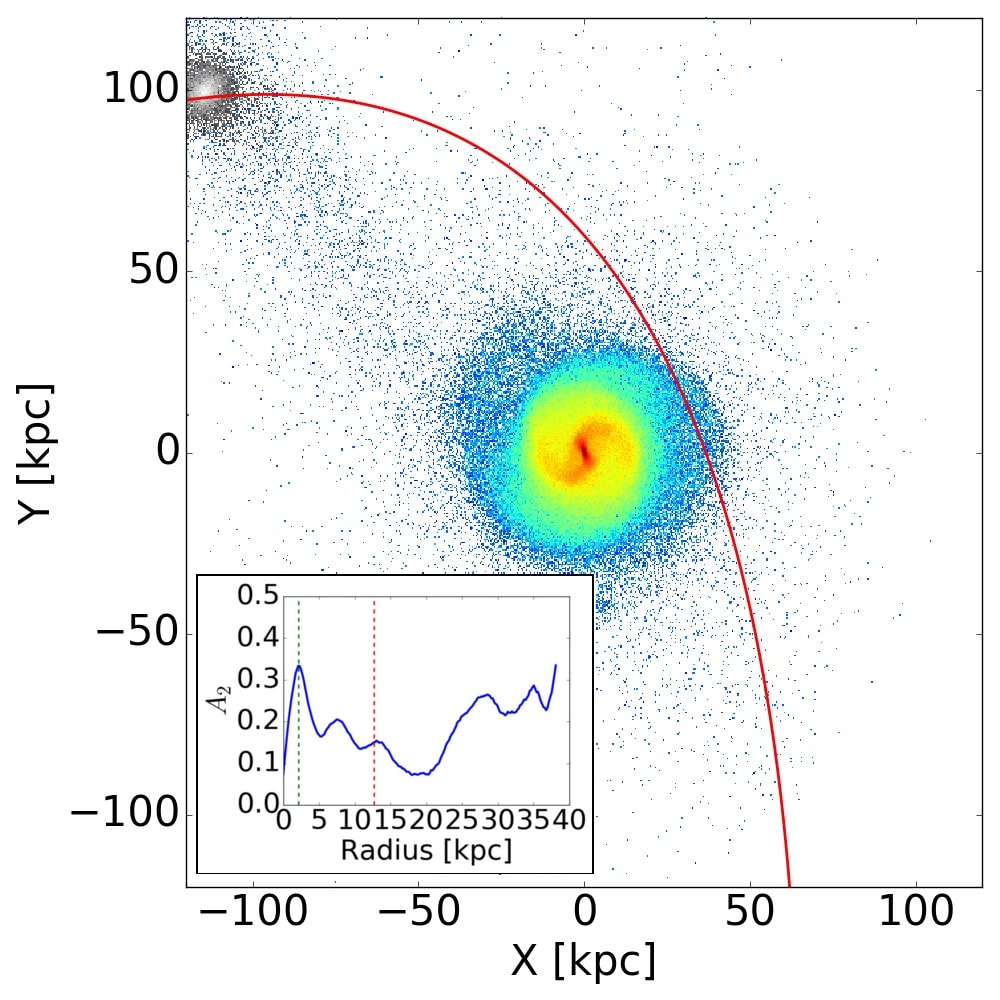}
\caption{Illustrations of six cases from our sample of interactions affecting the bar region (figure continued on the
next page). Each row corresponds to a different galaxy and each column is a different snapshot. In a given row one can
follow the interaction over time around the pericentre, from left to right. The galaxy in color scale is the primary
galaxy we study and the galaxy in grey scale is the perturber. The interpolated orbit of the perturber (see section
\ref{sample}) is plotted with a red line and the sense of rotation of the primary galaxy as a red arrow. The
coordinates $XY$ always represent the plane of the primary galaxy. Subpanels with the $A_2$ profile as a
function of radius are displayed, with the maximum of $A_2$ indicated with a dotted vertical green line and the
half-mass radius with a dotted vertical red line. In case there is no bar, the maximum of $A_2$ is at the half-mass radius. In the first case a zoom of the primary galaxy is displayed in the
top left corner of each panel.}
\label{examples}
\end{figure*}

\begin{figure*}
  \ContinuedFloat

    \includegraphics[scale=0.16]{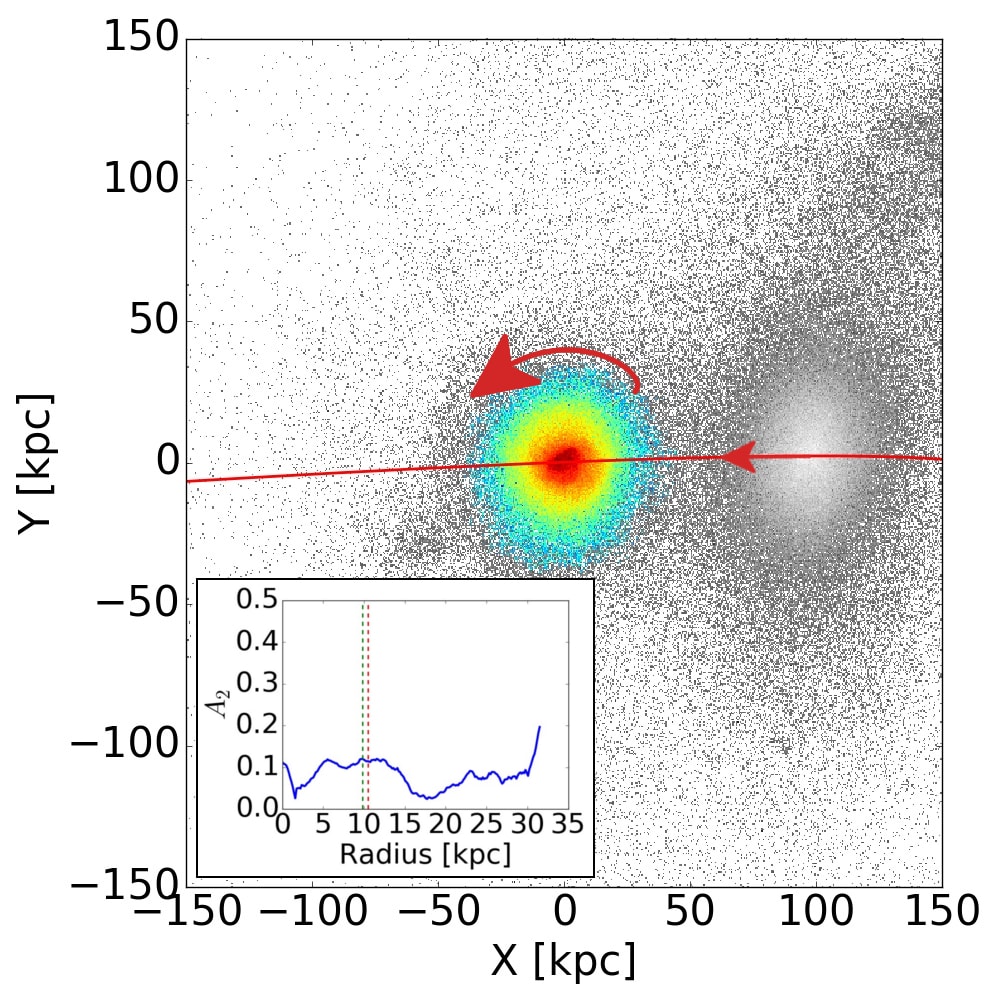}
    \includegraphics[scale=0.16]{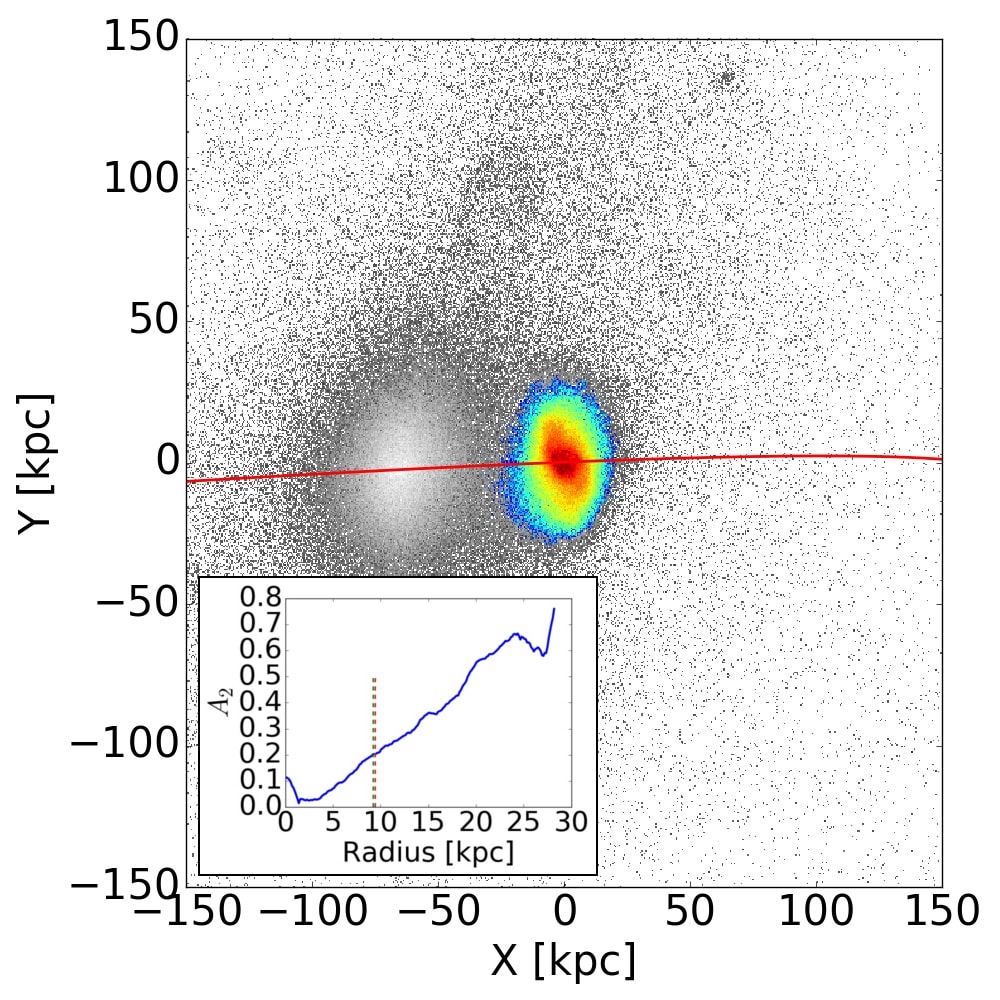}
    \includegraphics[scale=0.16]{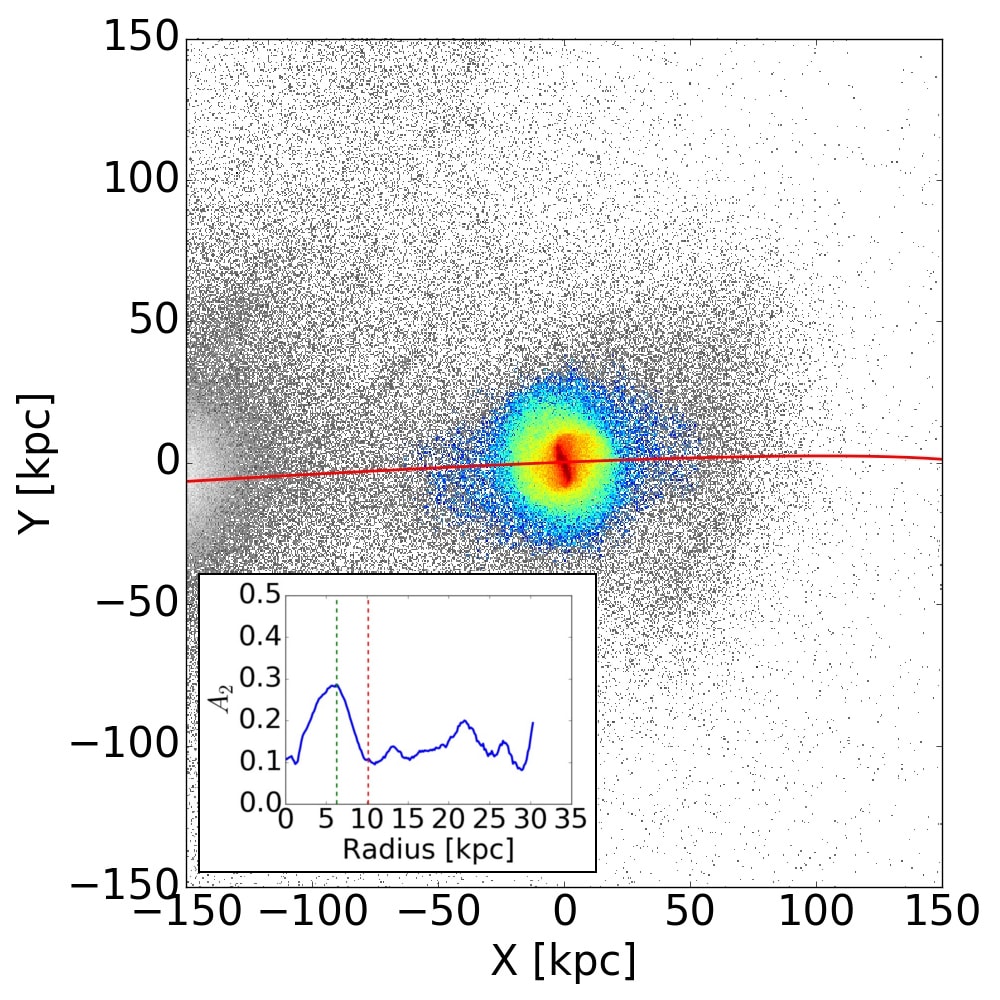}

    \includegraphics[scale=0.16]{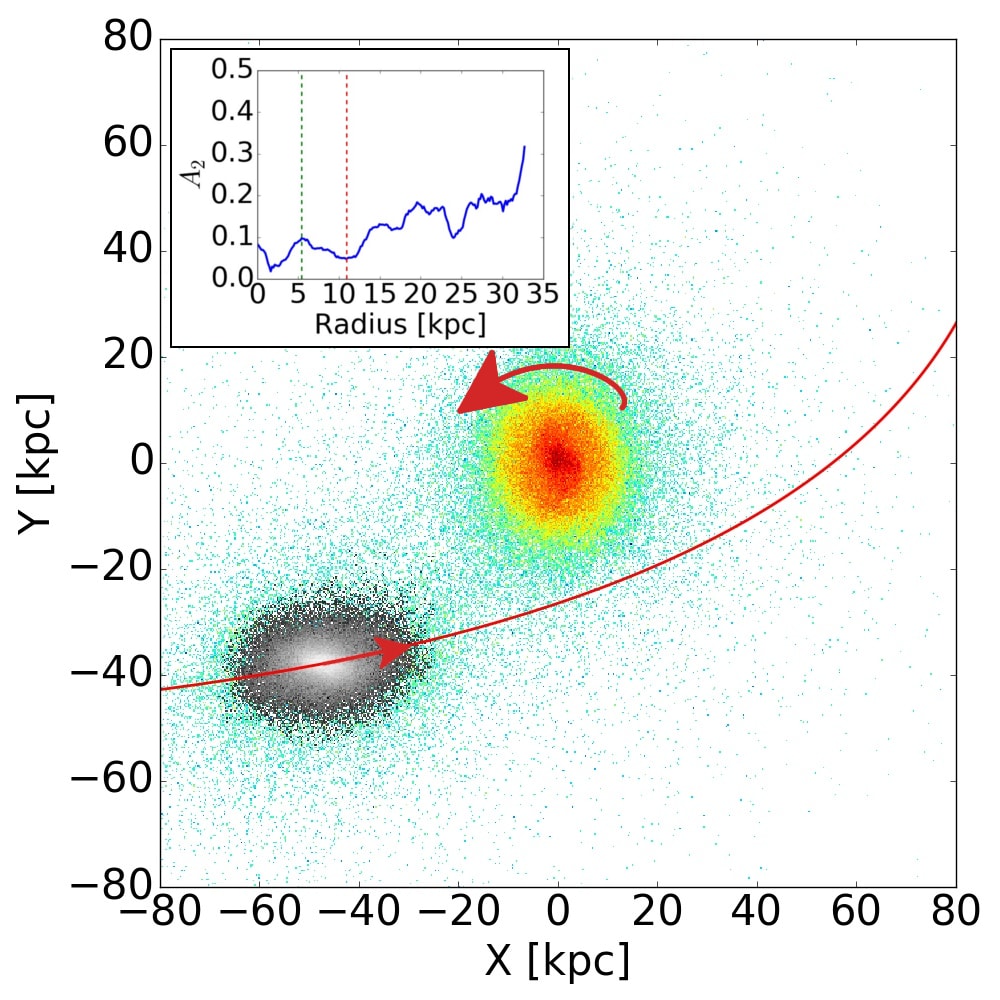}
    \includegraphics[scale=0.16]{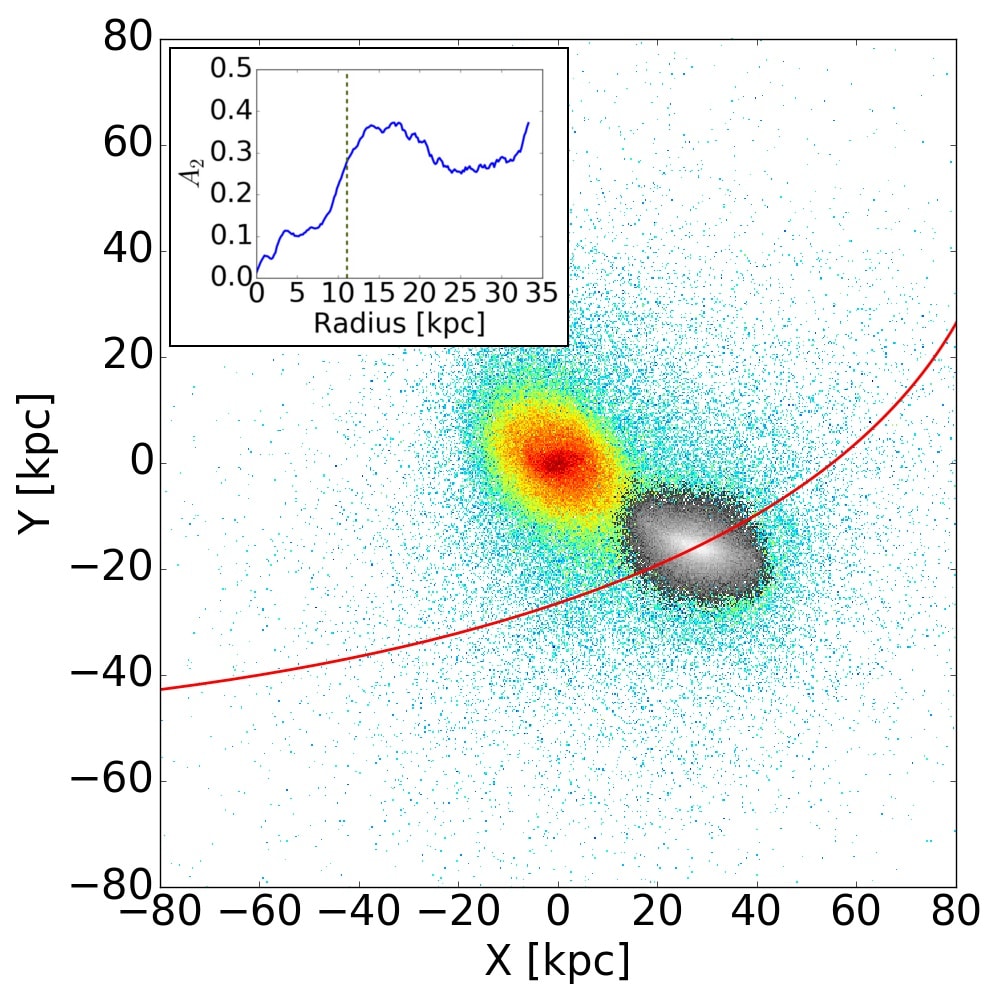}
    \includegraphics[scale=0.16]{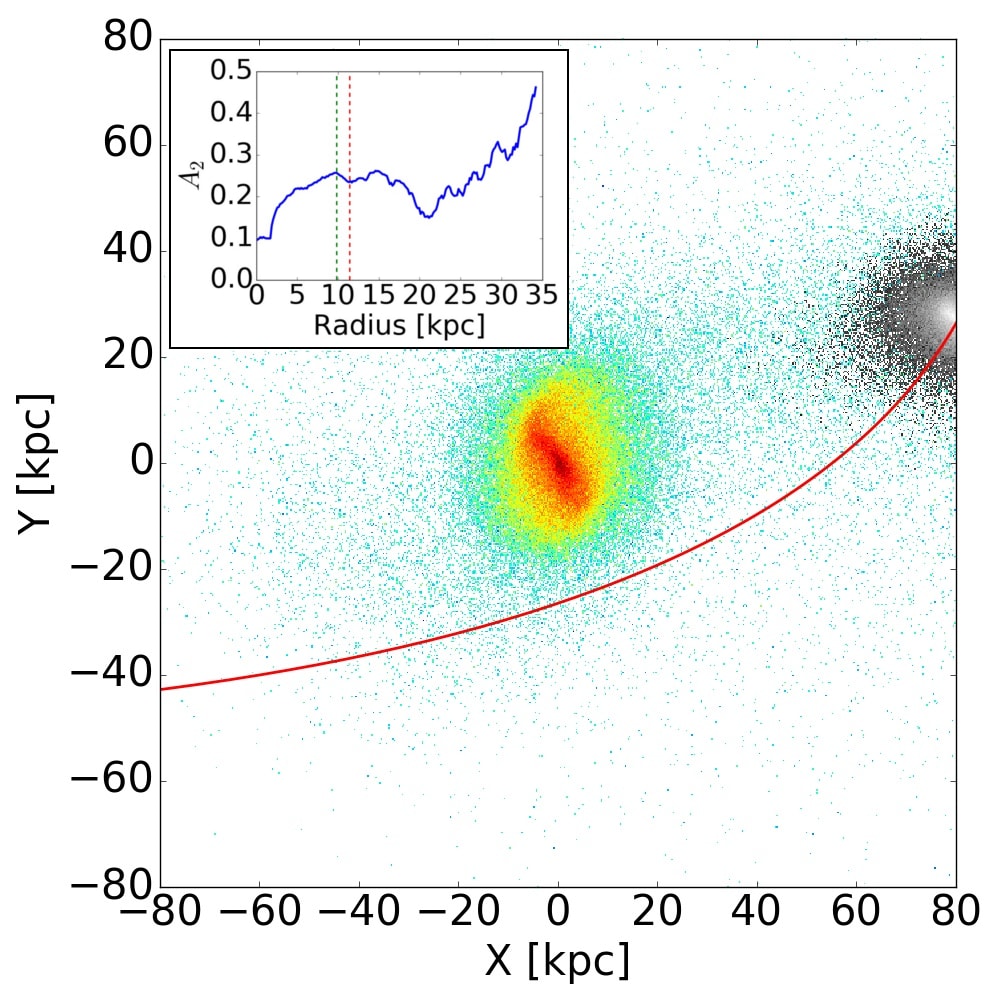}

    \includegraphics[scale=0.16]{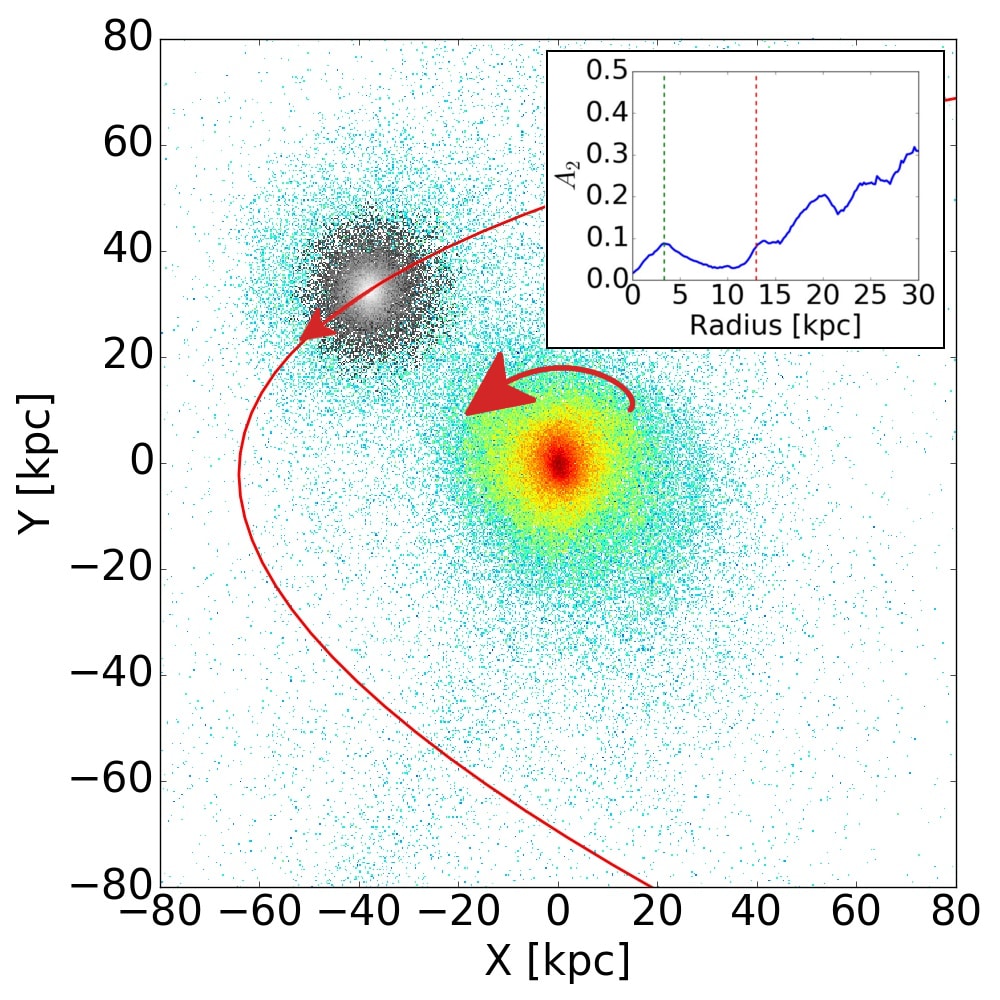}
    \includegraphics[scale=0.16]{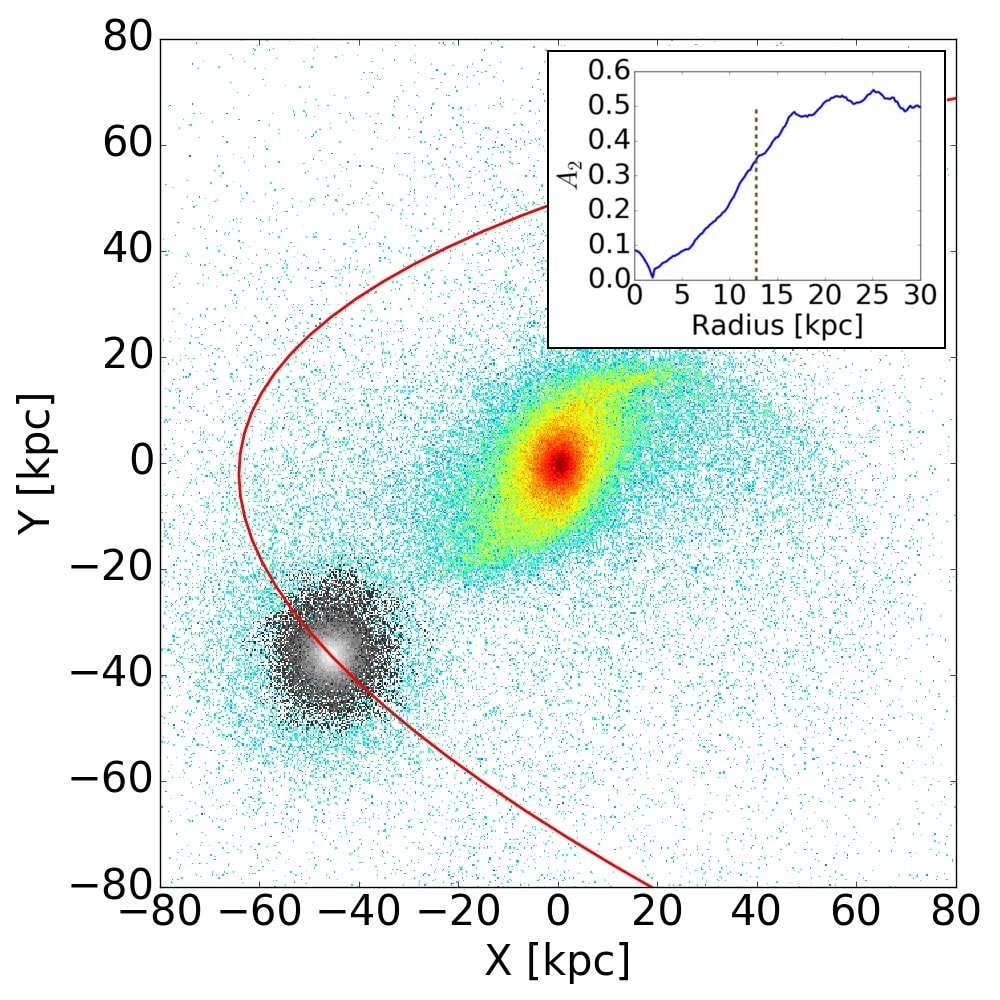}
    \includegraphics[scale=0.16]{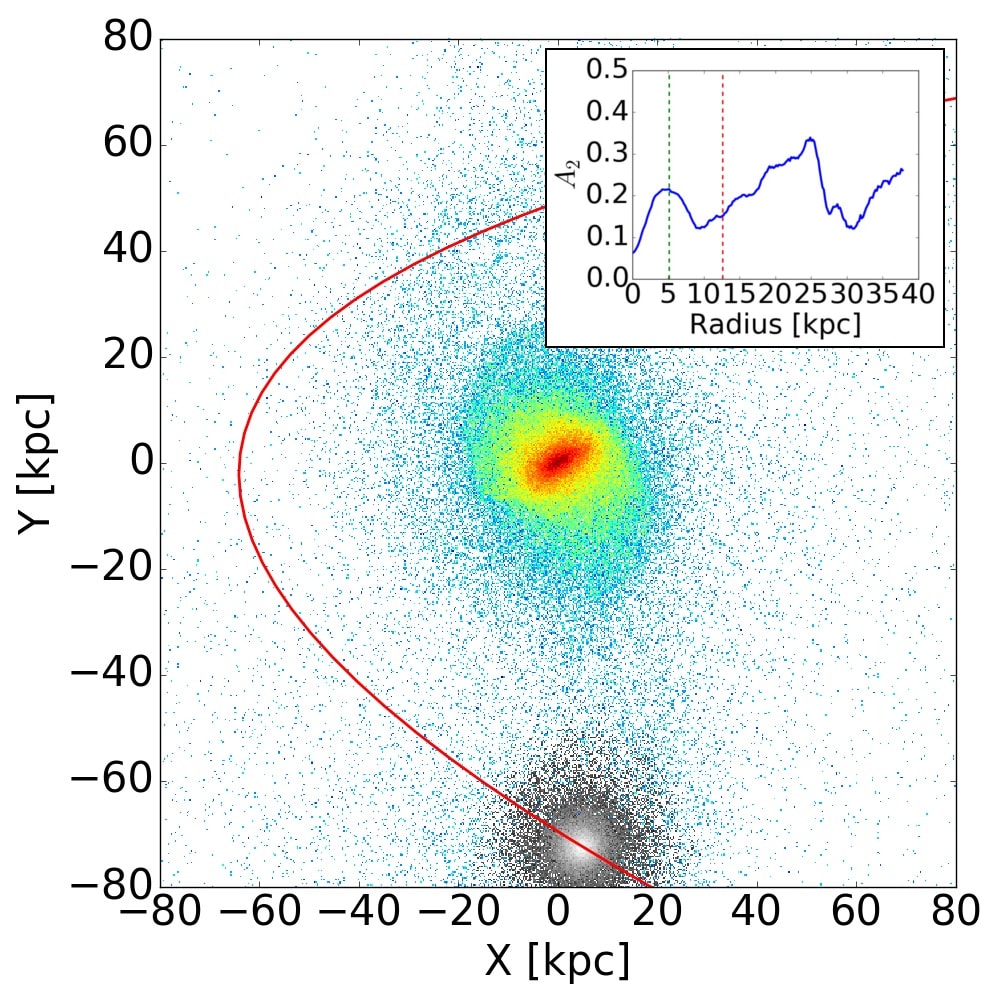}
    \caption{(continued)}
\end{figure*}


We will now focus on the cases of flyby events and their effect on the bar. As in section \ref{origin}, we start
with barred galaxies at redshift $z=0$ and look backward in time. This time however we do not restrict ourselves to
the moment of bar creation but look at every flyby interaction that affects the bar strength in any way. We keep
looking backward in time until the bar does not exist any more or the number of stellar particles drops below 40
000 in the galaxy. We define a flyby interaction as the passage of another galaxy (called perturber) close to the
primary galaxy, with no immediate merger involved. A merger can follow after the apocentre passage, but we keep only
cases where there are at least four snapshots between the first pericentre and the moment of the merging, to be able to
isolate the effect of the flyby event on the central region corresponding to the bar, before the merger.

We retain only the clear cases where a flyby creates a bar or impacts the bar strength in the case of a pre-existing
bar. In the latter case, we look for a change of at least 10\% in the bar strength. We discard all the cases (probed by
eye) where several interactions come into play simultaneously or when the bar strength variation may not be due to the
considered flyby. In total, this gives us a sample of 121 interaction cases affecting the bar strength, 50 of which are
bar creations. We call this sample the \textit{interaction sample}. A summary of the different samples is
presented in Table~\ref{table}.

In Fig.~\ref{examples} we present six examples from our sample to show how the
interactions can trigger or affect the bar in Illustris galaxies. We display in each case the evolution over three
snapshots of the primary galaxy and its perturber: one snapshot before the pericentre, the snapshot closest to the
pericentre, and one snapshot after the pericentre. This allows us to visualize how an initially unbarred (or weekly
barred) disc develops a (stronger) bar after the pericentre passage, due to the tidal forces induced by the perturber.
In the subpanels of Fig.~\ref{examples}, we plot the corresponding radial $A_2$ profiles of the primary galaxy
in these six examples. One can see how each disc develops a peak in the central region of the $A_2$ profile,
corresponding to the bar.

As we often have only a few snapshots showing each of the flyby events, it is not straightforward to determine the
exact time of the pericentre, as well as the pericentre distance. We thus decided to interpolate the orbit of the
perturber galaxy around the primary, using three snapshots in the vicinity of the pericentre. This interpolation was
done using a parabola with free parameters in the orbital plane (defined from the three subsequent positions of the
perturber). Although simplistic, this approach gives us a first approximation of the pericentre and realistic
orbits. Furthermore, we interpolated the velocity of the perturber along the orbit. We assumed the velocity is in the
orbital plane, and interpolated the $x$ and $y$ components (in the plane) separately, using again a parabola based on
the three snapshot values. These velocities allowed us to derive the precise time of the pericentre.

In our interaction sample, there is a number of cases where the perturber does not go around the primary galaxy but
goes through the primary disc before going out. In this situation, the tidal forces are not the only ones coming into
play, as the collision can directly disrupt the disc and affect the bar region. We call these interactions
\textit{fly-through} cases, to distinguish them from the real \textit{flybys} in our interaction sample. To
separate these two sub-samples, we use the interpolated orbits, in particular the interpolated pericentre distance.
We use the stellar half-mass radii of both the primary $R_{h,1}$ and the perturber galaxy $R_{h,2}$ and define
fly-throughs as the cases where the pericentre distance is lower than the sum $R_{h,1}+R_{h,2}$, i.e. when both galaxies
overlap at the pericentre. This gives us 79 cases of flybys and 42 cases of fly-throughs in our sample.

We also construct a reference sample, containing only those interacting cases that do not seem to affect the bar in any
way. By that we mean interactions during which the bar strength of the primary galaxy remains approximately constant
(with variations smaller than 10\%, and after visual inspection).
To build this sample, we proceed as previously, i.e. we follow disc galaxies
hosting a bar backward in time from redshift $z=0$ until the moment of bar formation and look for interactions
occurring while the bar strength stays constant. This is not obvious as the bar strength rarely remains constant for a
long time; it can vary due to the many external perturbations but also undergoes secular evolution over time. We
nevertheless found 70 cases of interactions not affecting the bar, which we will now refer to as
\textit{the reference sample}.

\subsection{Sample characteristics}
\label{sample_cha}

\begin{figure}
  \includegraphics[scale=0.3]{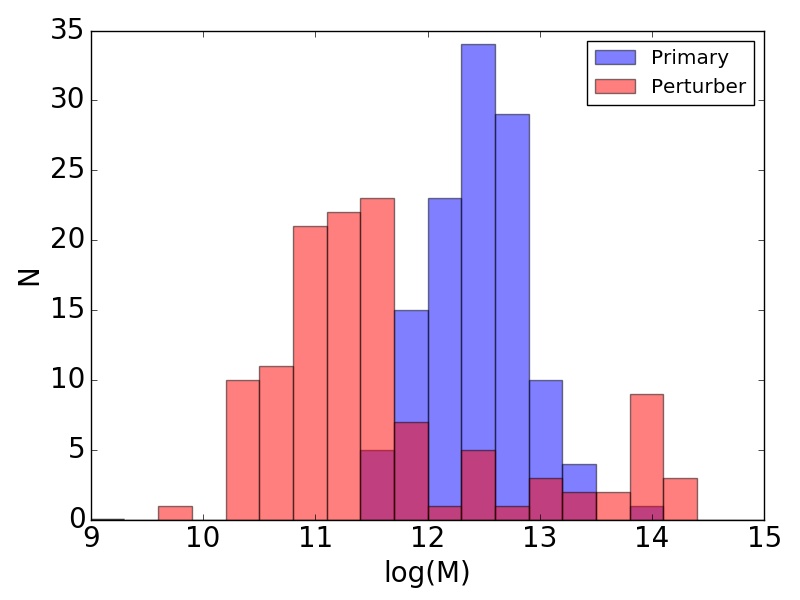}
\caption{Histograms of the total masses of the galaxies involved in the interactions of our flyby sample, i.e. for the
primary galaxies (in blue) and for the perturbers (in red).}
  \label{hist_m}
\end{figure}

\begin{figure}
  \includegraphics[scale=0.3]{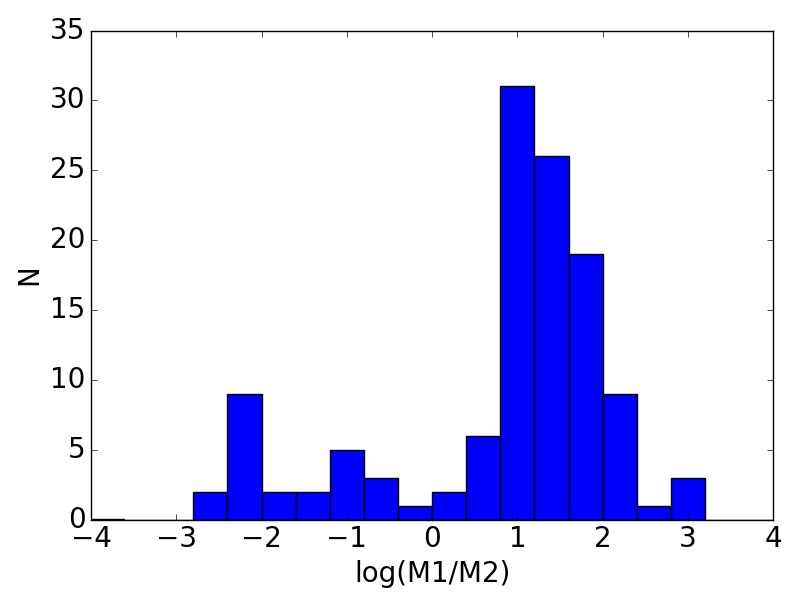}
\caption{Histogram of the total mass ratios (in log) of the galaxies involved in the interactions of our flyby sample.
$M1$ represents the masses of the primary galaxies and $M2$ of the perturbers.}
  \label{hist_mratio}
\end{figure}

\begin{figure*}
  \includegraphics[scale=0.3]{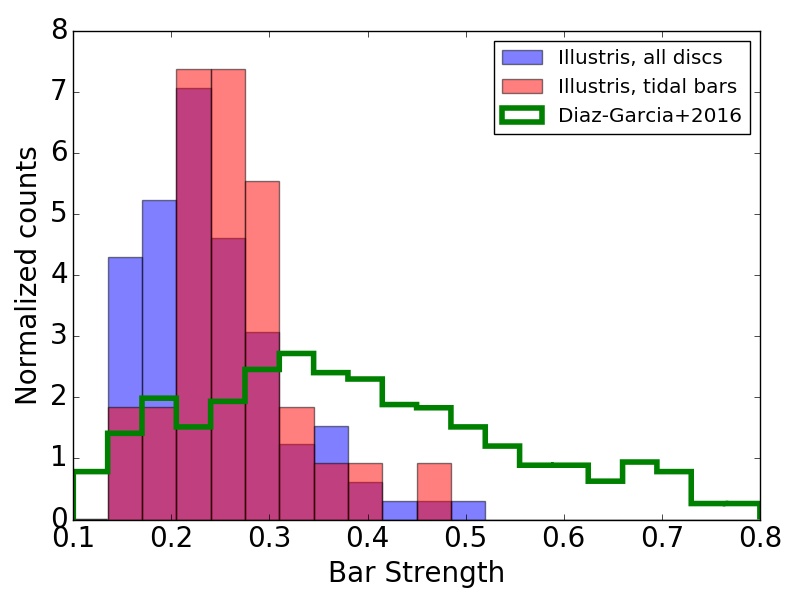}
  \includegraphics[scale=0.3]{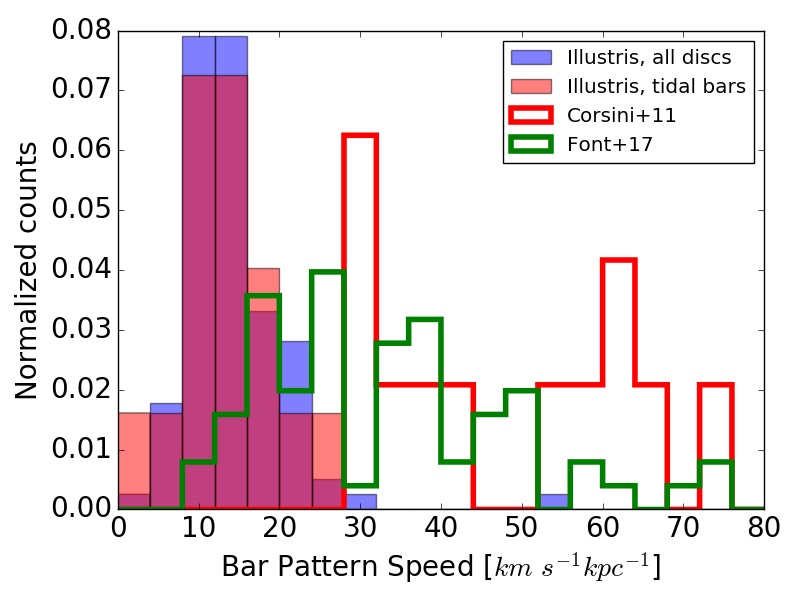}
  \includegraphics[scale=0.3]{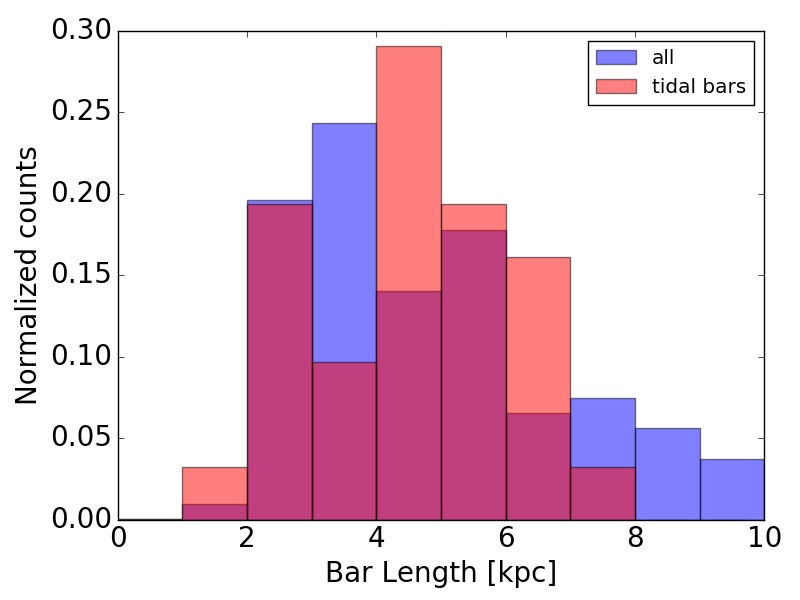}
  \includegraphics[scale=0.3]{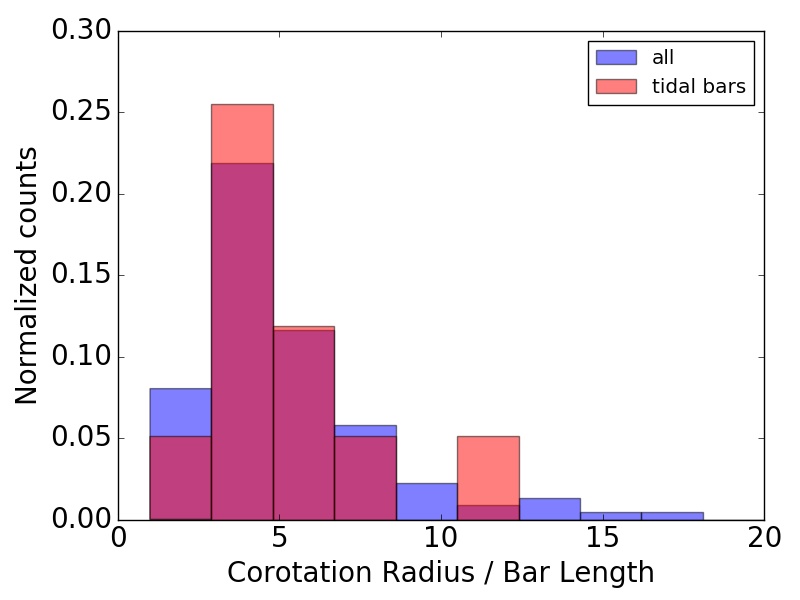}
\caption{Normalized histograms of the bar strengths (top left panel), bar pattern speeds (top right panel), bar
lengths (bottom left panel), and the ratios of the corotation radius and the bar length (bottom right panel),
for our flyby sample (in red, the values taken right after the interaction) and for all bars at redshift $z=0$
(in blue). For the strengths and pattern speeds we added observational values from
\citet{2016A&A...587A.160D}, \citet{2011MSAIS..18...23C}, and \citet{2017ApJ...835..279F}. Note that the sample of
all $z=0$ bars does not include the bars from our flyby sample, to allow a better comparison. The normalization is such
that the integral of the histogram is equal to 1. For the flyby sample we kept only the interactions happening
close to the present time ($z<0.21$) to be able to compare to the bars at redshift $z=0$.}
\label{hists}
\end{figure*}

\begin{figure*}
  \includegraphics[scale=0.2]{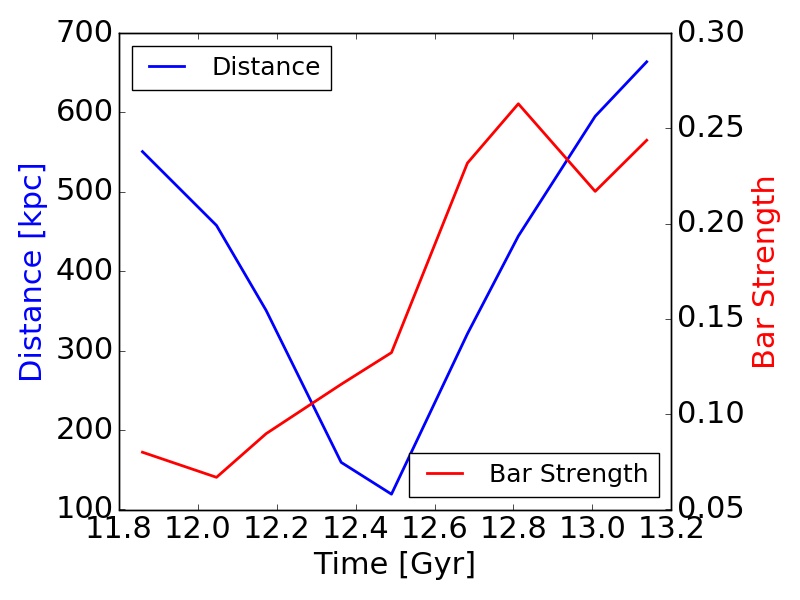}
   \includegraphics[scale=0.2]{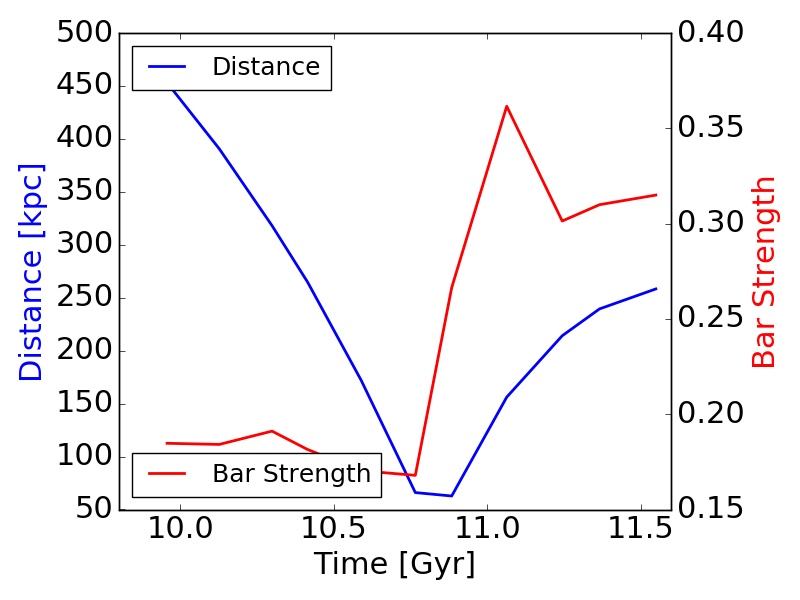}
  \includegraphics[scale=0.2]{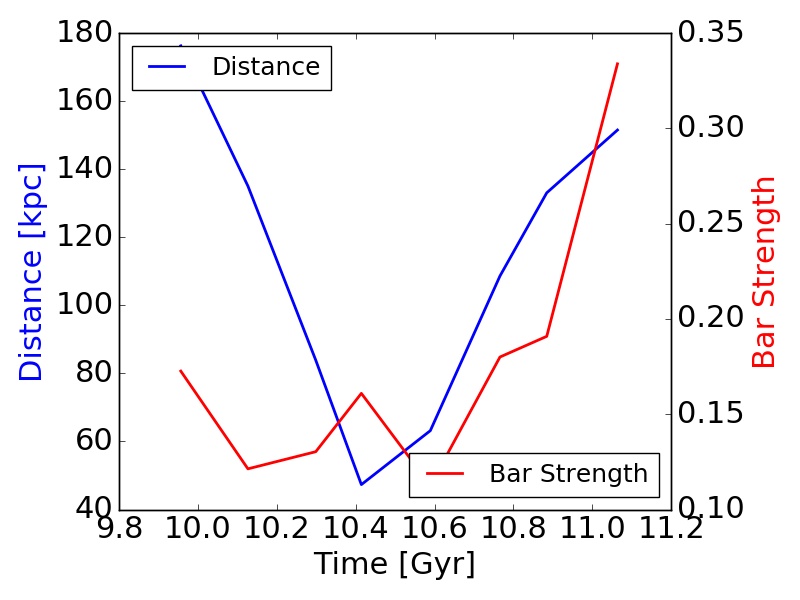}
  \includegraphics[scale=0.2]{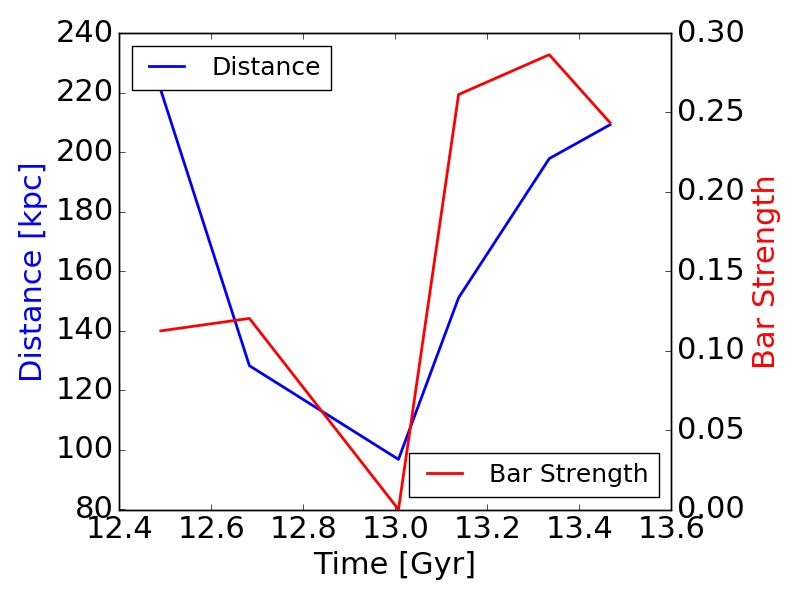}
  \includegraphics[scale=0.2]{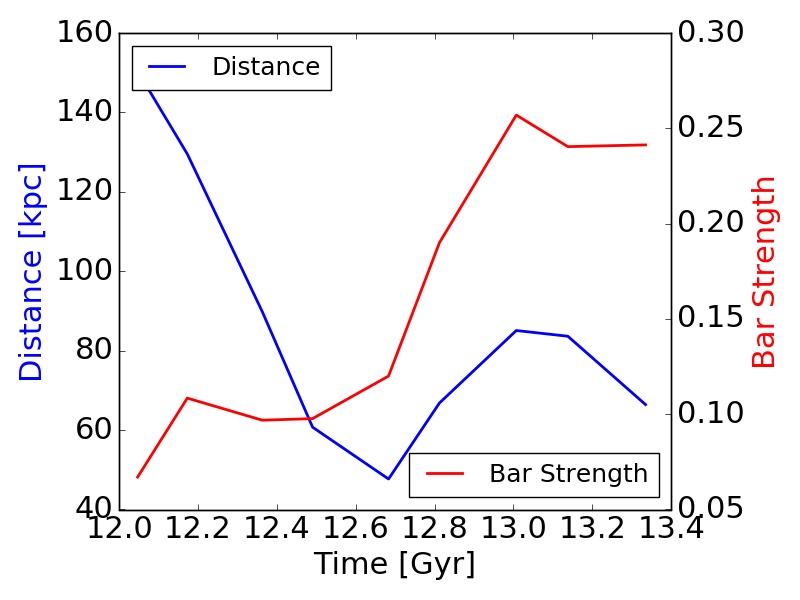}
  \includegraphics[scale=0.2]{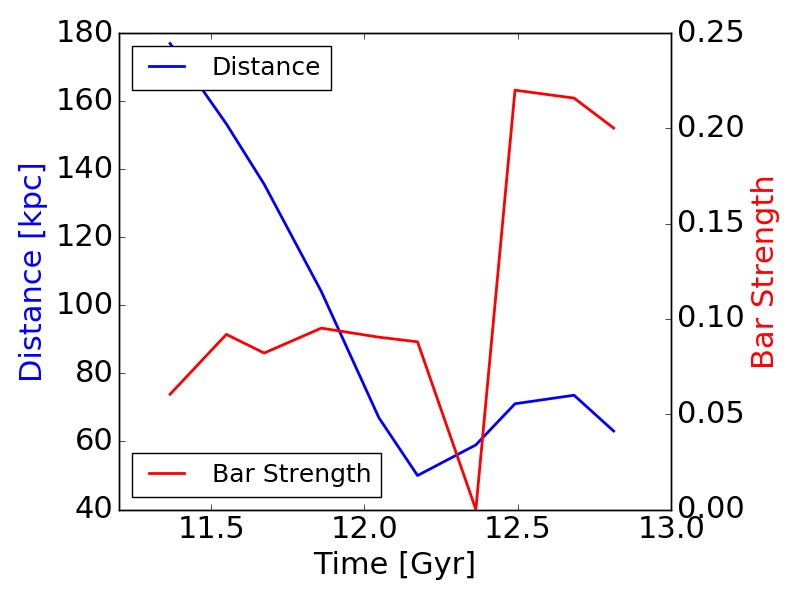}
  \caption{Bar strength (in red) and distance from the perturber (in blue) as a function of time, for the six cases
presented in Fig.~\ref{examples}. The increase in the bar strength occurs right after the pericentre
passage.}
  \label{dist}
\end{figure*}

We describe here in detail the characteristics of the interaction sample. The time of the interaction covers the range
from redshift $z=1.9$ (snapshot 69) to redshift $z=0$ (snapshot 135). The primary galaxy is not necessarily more
massive than the perturber in our interaction events; the primary can also be a satellite moving close to a bigger
galaxy (the perturber), but we still see it as a flyby (or fly-through) of the perturber, as we focus on the primary
disc. We compute the masses of the primary and the perturber galaxies, taken one snapshot before the pericentre, since
when two galaxies are too close, Illustris sometimes wrongly attributes particles to those galaxies. Furthermore, we
did not use the masses provided by Illustris, but instead fitted the dark matter density profile with an NFW profile to
derive its scale radius, and defined the total mass as the mass of all the particles inside ten times this scale
radius. Histograms of the total masses of both the primary and the perturber before the interaction are shown in
Fig.~\ref{hist_m}. One sees that the primary galaxies are mostly quite massive disc galaxies, while most perturbers
tend to have lower masses, with a few exceptions of huge elliptical galaxies. We also display the histogram of the mass
ratios between the primary and the perturber in Fig.~\ref{hist_mratio}. Out of 121 cases, only 24 have a perturber
bigger than the primary galaxy. This is probably due to the fact that since most our primary galaxies are massive, it
is not very common to see an interaction with an even more massive galaxy, while smaller perturbers are very numerous.

As far as the bars in our interaction sample are concerned, we are interested in comparing their strength, length and
pattern speed to non-perturbed bars. Since these quantities vary with redshift (see section~\ref{partI}), the
comparison has to be done for a given epoch. We thus look at all the galaxies in our interaction sample with a
pericentre passage occurring between redshift $z=0$ and $z=0.21$, and compare them to the sample of all barred galaxies
in our default sample (see section~\ref{barstr}) at redshift $z=0$. We obtain the values of the bar strength, bar
length and pattern speed after the pericentre in our interaction sample. We derive those values by averaging
them over between two and four snapshots after the pericentre. Indeed, in the snapshot right after the pericentre the
bar is often still in formation, and after its creation its strength can vary quickly over time so that it is safer to
average over a few snapshots.

For the bar strength, we find that the bars affected by interactions tend to be stronger on
average (see Fig.~\ref{hists}, top left panel) than the mean of all the bars at low redshift, with the mean value
around 0.26 $\pm$ 0.06, against 0.24 $\pm$ 0.07 in the total sample (where the bars from the interaction sample
have been removed). This result is consistent with previous studies showing that interactions tend to produce
stronger bars than bars formed secularly (\citealt{1998ApJ...499..149M}; \citealt{2004MNRAS.347..220B};
\citealt{2014MNRAS.445.1339L, 2016ApJ...826..227L}). Note that if we take the standard error on the mean instead of the standard deviation, we find errorbars of respectively 0.01 and 0.008, so that our mean values seem to be distinct in both samples. Our bars also tend to be slightly slower in the
interaction sample than in the total sample (see Fig.~\ref{hists}, top right panel), with a mean value of the pattern
speed $13.5 \pm 5.8$ km s$^{-1}$ kpc$^{-1}$, against $14.2 \pm 6.4$ km s$^{-1}$ kpc$^{-1}$. We find standard errors on the mean to be here respectively 1.0 and 0.6 km s$^{-1}$ kpc$^{-1}$. Therefore, the
uncertainties are quite large, and the difference in the mean values is quite small, so the values for both samples
remain compatible.

We also tried to probe how fast or slow our bars are in terms of the ratio of the corotation radius to
the bar length, $\mathcal{R}$ (e.g. \citealt{2000ApJ...543..704D}). We derive the
corotation radius by computing the circular velocity as a function of radius $R$: $v_{\rm circ}(R)=(GM/R)^{1/2}$, where
$M$ is the total mass inside radius $R$.
We then look for the intersection between the circular velocity and the bar
velocity $v_{\rm bar}(R)=R\Omega_p$ to derive the corotation radii, which we divide by our bar lengths to obtain
$\mathcal{R}$.
As our bar lengths are not directly
comparable to observations, neither are our values of $\mathcal{R}$, but we can compare the values between the
interaction sample and the total discs sample (see Fig.~\ref{hists}, bottom right panel). With this definition, we find
the bars in the interaction sample to be slightly faster than in the total sample, with a mean value of 5.6 $\pm$ 3.8 (error on the mean 0.3) for the
total sample, and 5.1 $\pm$ 2.6 (error on the mean 0.5) for the interaction sample. This is in contradiction with our bars having a lower
pattern speed in the interaction sample, which might be due to our corotation radii being not reliable, or to the
choice of defining the bar length as the maximum of the $A_2$ profile.


As far as the bar length is concerned, we do not see a significant difference between the interaction sample
and all bars at low redshift (see Fig.~\ref{hists}, bottom left panel), with the mean value of the bar length $4.7
\pm 2.0$ kpc (error on the mean 0.2) in the total sample, i.e. slightly higher than in the interaction sample ($4.5 \pm 1.5$ kpc, error on the mean 0.3). Here
again the difference is nevertheless small compared to the uncertainties.

\subsection{Tidally induced bars}
\label{tidal}

\subsubsection{Effect of the interaction on the bar region}

There are two different scenarios for the tidal bars of our interactions sample: either there was already a bar in the
primary galaxy before the passage of the perturber, or the galaxy was at first unbarred. In the latter case, it means
that the interaction created the bar, while the former case corresponds to interactions affecting a pre-existing bar.
The passage of a perturber can either reinforce the bar (bar strength increases), or weaken it (strength decreases). We
thus have three subsamples: bar creating interactions (50 cases), bar increasing interactions (53 cases) and bar
decreasing ones (18 cases).
In Fig.~\ref{examples}, the second example is a case of bar increase, with a very weak bar in the centre being
reinforced by the interaction. All the other cases shown in this Figure are bar creating interactions, as there is no
clear bar in the centre before the interaction.

\begin{figure}
  \includegraphics[scale=0.3]{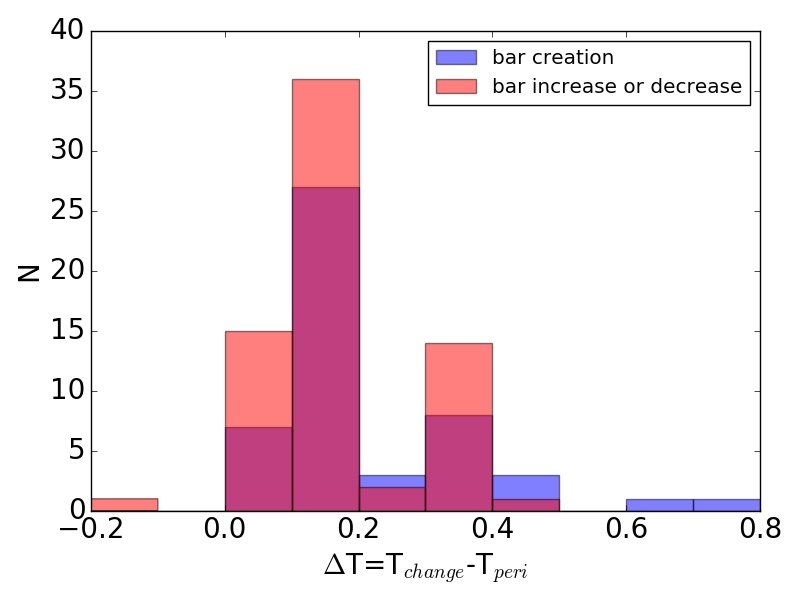}
\caption{Histogram of the difference between the time of change in the bar strength and the pericentre time of
the interaction, in our sample of 121 interactions affecting the bar. The blue histogram is for interactions
creating a bar and the red one for interactions affecting a pre-existing bar.}
  \label{tcorr}
\end{figure}

To look at the effect of the interaction in more detail, we examine the bar strength as a function of time, as shown in
Fig.~\ref{dist} for our six examples from Fig.~\ref{examples}. We find that the change in the bar strength (whether it is an increase or a decrease) seems to
happen right after the pericentre passage, the bar strength being more or less constant before. To characterize this,
we find in each case the snapshot at which the bar strength changes: for the bar creating cases, we take the time
when it becomes higher than 0.15, while for the bar increasing and decreasing cases, we take the maximum of the bar
strength derivative. Plotting this time minus the time of the pericentre (Fig.~\ref{tcorr}), we confirm that the bar
strength increase happens right after the pericentre, on average 0.18
$\pm$ 0.14 Gyr after (0.21 $\pm$ 0.16 Gyr for bar creation, 0.17 $\pm$ 0.12 Gyr for bar increase and decrease).

We will now try to understand what causes the bar strength to increase or decrease in our sample. In the following
subsections, we look at two parameters, the orbital angle and the strength of the interaction, to see if they are
responsible for the way the bar is affected by the interaction.

\subsubsection{Orbital angle}

We define the orbital angle of the interaction as the angle between the primary disc plane and the plane of the
perturber's orbit, i.e. between the primary disc rotation axis and the angular momentum $L_z$ of the perturber in the
primary galaxy frame. This gives us an angle between 0 and 180 degrees, with 0 corresponding to a prograde interaction
(orbit of the perturber in the disc plane with the disc rotating in the same sense as the perturber moves on its orbit)
and 180 to the retrograde case (orbit of the perturber in the disc plane with the perturber going in the opposite
sense to the disc rotation). This angle is only relevant for flyby events, as for a fly-through the perturber goes
through the primary disc instead of orbiting around. Therefore in this section we restrict ourselves to flyby cases,
i.e. 79 interactions.

We derived this angle averaged around the pericentre passage for our sample of 79 flyby interactions and took the cosine of
the angle to get a normalized distribution, since the possibilities of the perturbers's direction with respect to the
primary disc form a solid angle. We plot in Fig.~\ref{angle} (right panel) the histogram of the orbital angles (in
cosine) in our sample, for the bar creating and bar increasing interactions only. We find that even though the
interactions span the whole range of values of the orbital angle, most interactions creating a bar or increasing the
bar strength are at low orbital angles ($\cos(\theta) \sim 1$), i.e. correspond to more or less prograde cases.
Therefore, prograde interactions seem to be the preferred way of creating a bar or increasing the bar strength of a
pre-existing bar.

The fact that a prograde orbit of the perturber is more efficient to create or increase a bar is in agreement with
previous work (\citealt{1990A&A...230...37G}; \citealt{2008ApJ...687L..13R}; \citealt{2014ApJ...790L..33L}; \citealt{2015ApJ...810..100L}; \citealt{2017ApJ...842...56G};
\citealt{2018arXiv180309465L}), and can be easily understood by looking at the tidal force acting on the primary galaxy
during the interaction. In a prograde case, the perturber is moving around the primary galaxy while the latter is
rotating in the same sense, so that the parts of the primary most affected by the tidal force remain approximately the
same over time. Therefore these parts will be very elongated by the tidal force. On the contrary, for a retrograde
orbit, the parts most affected by the tidal force will change continuously during the passage of the perturber as the
primary galaxy is rotating in the opposite direction, so that the elongated parts will not be consistent over time, and
therefore the bar shaped elongation will be weak.
Although a retrograde orbit can still affect a bar, it will be much less effective in doing so.

\subsubsection{Strength of the interaction}
\label{strint}

Another parameter to take into account is the strength of the interaction. A small satellite at a large distance is
expected to have less impact on the bar region than a big perturber close to the primary disc. To characterize this, we
use the Elmegreen tidal strength parameter $S$ (\citealt{1991A&A...244...52E}), defined as:
\begin{equation}
 S=\frac{M_{\rm pert}}{M_{\rm gal}} \left(\frac{R_{\rm gal}}{R_{\rm peri}}\right)^3 \frac{\Delta T}{T}
\end{equation}
where $M_{\rm pert}$
and $M_{\rm gal}$ are the perturber and primary galaxy total masses, while $R_{\rm gal}$ and $R_{\rm peri}$ are the
galaxy size (taken here as eight times the stellar half-mass radius) and the pericentre distance, respectively. $\Delta T$ is
the time it takes the perturber to travel one radian around the primary galaxy centre near the pericentre. To
compute it, we use the orbit interpolation described in section~\ref{sample}, as well as the velocity interpolation of
the perturber. $T$ is the time it takes the stars in the outer parts of the primary disc to travel one radian
around the centre and can be derived with:
\begin{equation}
 T=\left(\frac{R_{\rm gal}^3}{G M_{\rm gal}}\right)^{1/2}.
\end{equation}
The interaction is thus stronger, as expected, when the perturber is more massive than the primary galaxy, the
pericentre distance is small, and the perturber has a low velocity with respect to the primary galaxy.

\begin{figure}
  \includegraphics[scale=0.25]{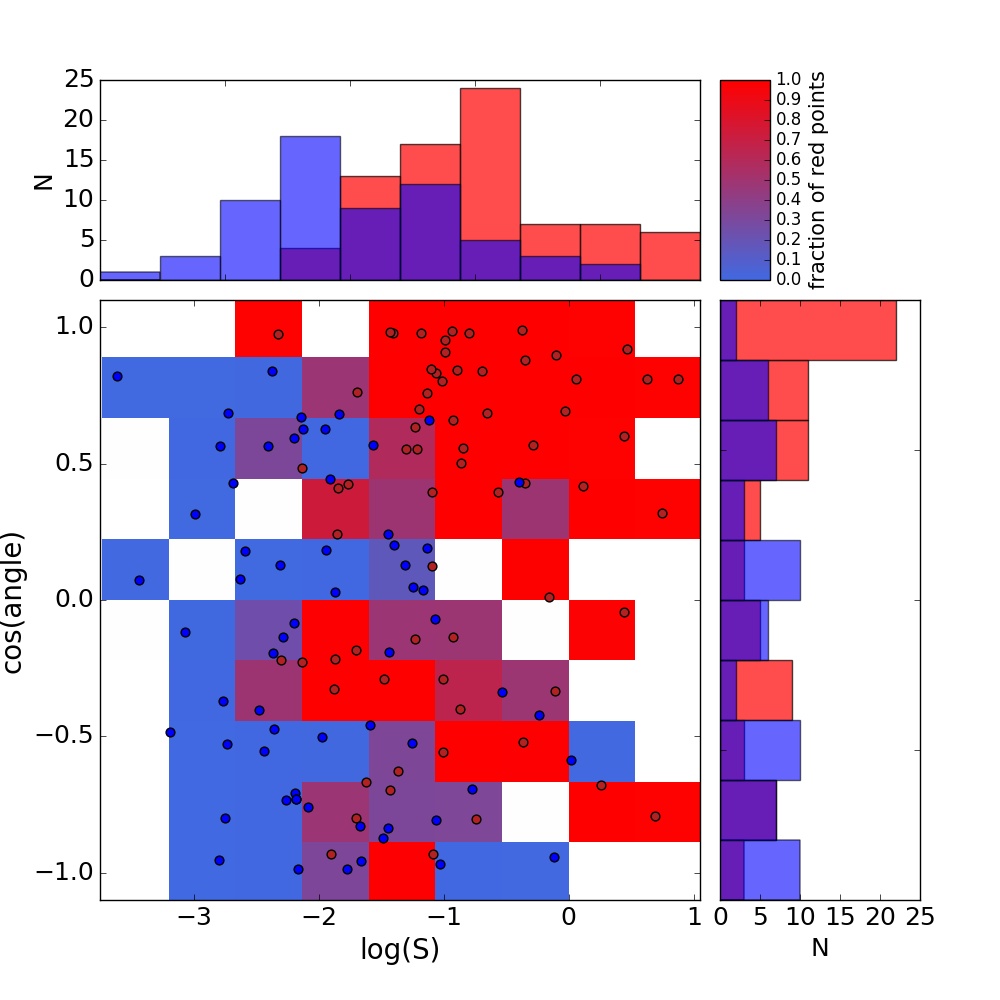}
\caption{Top panel: histograms of the Elmegreen tidal strengths of interactions (in log), for the flyby interaction sample (79 cases) of bar creating and bar reinforcing cases
(in red) and for the reference sample (i.e. interactions not affecting the bar, in blue). Right panel: histograms of
the orbital angles (in cosine) for those two samples. Middle panel: orbital angles as a function of the Elmegreen tidal
strengths for these two samples, with red points for the flyby interaction sample and blue points for the reference sample. We added a color scale to
show whether the regions are dominated by the interaction sample or the reference sample: each bin is colored by the
fraction of red points in it.}
\label{angle}
\end{figure}

We derived the tidal strength parameter for our sample of interactions, using the total masses derived in section~\ref{sample_cha}, and using the half-mass radius as derived from an exponential fit of the stellar density profile for the primary galaxy.
We compare the values obtained for the 79 flyby cases to the ones of the reference sample (see section \ref{sample}) in Fig.~\ref{angle}. We find that,
as expected, strong interactions tend to systematically affect the bar, so that the sample of interactions without
effect on the bar have lower Elmegreen parameters. Furthermore, the top right part of the main plot (middle panel) in
Fig.~\ref{angle} is mostly red, showing that a strong, prograde interaction is very likely to affect the strength of
the bar, whereas a retrograde weak interaction is not.

\begin{figure}
  \includegraphics[scale=0.3]{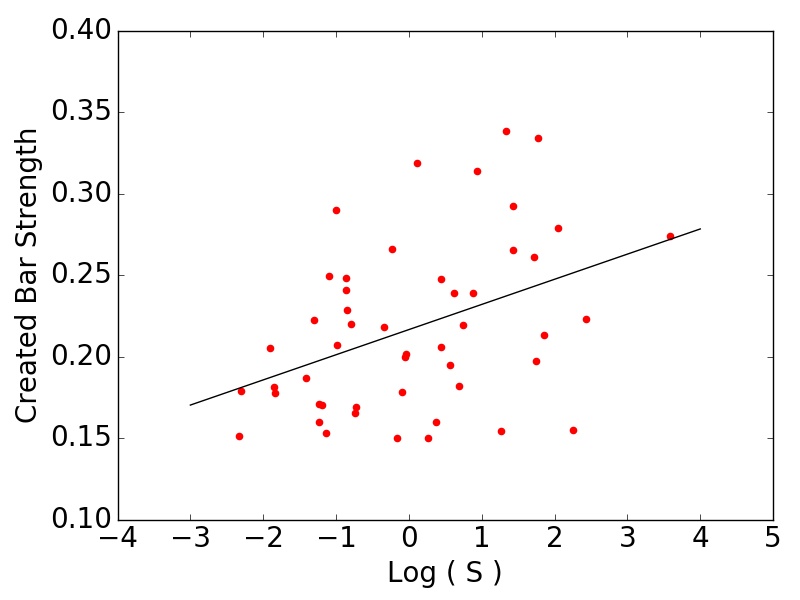}
\caption{Strength of bars created by a tidal interaction in our interaction sample (50 cases), as a function of the Elmegreen tidal strength of the
interactions (in log). The black line represents the linear regression, with a correlation coefficient of $r^2$=0.17
and a $p$-value of $3.56 \times 10^{-3}$.}
\label{corr}
\end{figure}

Now we would like to check if the strength of the interaction can be linked to the resulting bar properties. In
particular, are the bars obtained after a strong interaction stronger? We take the bar creating interactions subsample (50 cases, including both flybys and fly-throughs),
and look at the strength of the bar created by the interaction. As the bar strength is evolving after the interaction,
we average the bar strength over two snapshots after the pericentre. In Fig.~\ref{corr} we plot for every bar
creating interaction of our sample the derived bar strength as a function of the tidal strength parameter. We find a
correlation coefficient of 0.17 and a $p$-value of 0.0036, showing that stronger interactions do tend to create
stronger bars, but with a weak trend. Stronger bars forming as a result of stronger tidal forces are consistent with
previous studies (\citealt{2004MNRAS.347..220B}; \citealt{2016ApJ...826..227L}). The large scatter observed in our plot
can be explained by the fact that other parameters can come into play in determining the strength of the created bar.
In particular, the orbital angle of the perturber, the primary disc stability or the gaseous content of the primary
disc, all can directly affect the creation of the bar and its strength.

We found no correlation of the bar strength increase with the Elmegreen tidal strength parameter for the bar increasing
subsample, which might be due to the fact that in addition to all these parameters, the pre-existing bar position angle
with respect to the perturber position may determine whether the interaction will increase or decrease the bar strength
(\citealt{1990A&A...230...37G}; \citealt{2014MNRAS.445.1339L}). Unfortunately, our time resolution is not good enough to be able to accurately derive
the bar position angle at the precise time of the pericentre.

\section{Discussion}
\label{discuss}

The barred galaxies we have considered in this paper are in a high mass range ($M_*>8.3 \times 10^{10}M_{\odot}$),
since galaxies of lower masses do not seem to host enough bars (see section~\ref{barstr}). This mass range is higher
than in most observational studies looking at barred galaxies, which can make the comparison difficult, but Illustris
galaxies have been shown to be more massive than observed in the local Universe (\citealt{2017MNRAS.467.2879B}).
Furthermore, \citet{2018ApJ...853..194D} have recently shown that below stellar masses of $10^{11}M_{\odot}$, Illustris
galaxies seem to be unrealistic from the morphological point of view. This is in good agreement with our $8.3 \times
10^{10}M_{\odot}$ threshold, and it thus seems likely that below this value, galaxies are indeed not realistic enough
to produce bars.

In section \ref{barstr}, we used the threshold value of 0.15 in the bar strength to determine whether a galaxy is
barred or not (together with visual inspection). Other studies have used different values for this threshold, for
example \citet{2017MNRAS.469.1054A} who take it to be 0.2. Nevertheless, we have found in Illustris many galaxies with a
bar strength between 0.15 and 0.2 that visually clearly show a bar and thus decided to adopt 0.15 as sufficient.

The fact that in spite of restricting our study to high mass galaxies the bar fraction at redshift $z=0$ is quite low
(24\% at redshift $z=0$, while recent studies find bar fractions around 60\%, \citealt{2000AJ....119..536E};
\citealt{2009A&A...495..491A}; \citealt{2012MNRAS.423.1485S}; \citealt{2015ApJS..217...32B};
\citealt{2016A&A...596A..25L}; \citealt{2018MNRAS.474.5372E}), might be due to the gravitational softening scale. As
mentioned in section~\ref{origin}, it might be harder to form bars in secular evolution in Illustris, so that we expect
less bars overall. We verified that this low bar fraction is not related to the way we select galaxies for our
analysis, i.e. changing the flatness or the circularity constraints (see section~\ref{discs}) a little yields similar
bar fractions.

The only previous study of simulated barred galaxies in a cosmological context was published by
\citet{2017MNRAS.469.1054A}. They used the EAGLE Project (\citealt{2015MNRAS.446..521S}), which is a cosmological
hydrodynamical simulation based on GADGET3, simulating a 100 Mpc$^3$ volume containing nearly 7 billion particles.
Looking at the bar fraction at redshift $z=0$, we find significantly less bars than in the EAGLE simulation since
\citet{2017MNRAS.469.1054A} found about 40\% of barred galaxies. We also have weaker bars (see Fig.~\ref{frac_S}, as
well as Fig.~\ref{hists}, top left panel). This might point to a difference in the efficiency of bar formation between
Illustris and EAGLE.

Furthermore, the sample of galaxies considered by \citet{2017MNRAS.469.1054A} had stellar masses between
$10^{10.6}$ and $10^{11} M_{\odot}$, which in Illustris is too low to form bars consistently (see
section~\ref{barstr}). This may indicate that the reason why Illustris does not form bars for low mass galaxies is
linked to physical processes, such as baryonic physics. In particular, the AGN feedback in Illustris has been shown
to be excessive (\citealt{2014MNRAS.445..175G}). Our sample is therefore composed of more massive galaxies than
theirs.

Nevertheless, they find unbarred galaxies to be more gas-rich, in agreement
with our results. Furthermore, \citet{2017MNRAS.469.1054A} tend to have low pattern speed values in their strongly
barred galaxies at redshift $z=0$ (lower than 6 km s$^{-1}$ kpc$^{-1}$) than in our sample of galaxies at this time
(see Fig.~\ref{hists}, top right panel). Even if we restrain our sample to strong bars as they did, we still find higher
values with a mean over 10 km s$^{-1}$ kpc$^{-1}$. Note that the values we find for Illustris galaxies are more
consistent with other simulations of barred galaxies (both in isolation and tidally induced, e.g.
\citealt{2018arXiv180309465L}) as well as with observations (\citealt{2017ApJ...835..279F}, see Fig.~\ref{hists}).
We note however that Illustris bars cover only the lower end of the distribution of pattern speeds in observed
galaxies.

The presence of gas in the disc does not only tend to suppress bars formed in isolation, but also inhibits the creation
of tidally induced bars, as has been shown in \citet{2004MNRAS.347..220B}. Since most bars in Illustris seem to be
formed in interactions rather than in secular evolution, this could explain the low bar fractions we find in gas rich
discs, alongside with the fact that barred gas rich discs will in the long run weaken or even destroy the bar due to
gas inflows towards the centre.

In section \ref{barstr}, we found a trend of the bar fraction growing with the stellar mass of the galaxy. We also
studied the bar fraction as a function of the total galactic mass (including stars, gas and dark matter), but did
not find any clear trend. We looked as well for the dependence of the mean bar strength on the stellar mass, total mass
and the disc gas fraction, but again did not find any clear correlation between these properties.

In section~\ref{strint}, we used the Elmegreen parameter $S$ to characterize the strength of the interaction. This
parameter was originally designed to look at flyby interactions, however we also used it to characterize fly-through
events. Since the pericentre distance then becomes very small, this leads to $S$ taking very high values, as can be
seen in Fig.~\ref{corr}. Nevertheless, we still believe that this parameter is relevant to characterize the strength of
the interaction in this case, as fly-throughs are expected to be much more violent than flybys. We also tried to use
another definition of this strength, by computing the tidal force, defined as $R_{\rm gal}M_{\rm pert}/D^3$ (as in
\citealt{2016ApJ...826..227L}), where $R_{\rm gal}$ is the primary galaxy's size, $M_{\rm pert}$ is the perturber's
total mass, $D$ is the distance between the primary and the perturber. We then derived the integrated tidal force by
summing the tidal forces over three snapshots around the pericentre, using the interpolated orbit (see
section~\ref{sample}) and normalizing by the time that passed between the first and the last of the three snapshots. We
found a weak increasing trend of the created bar strength with the integrated tidal force, similar to the one in
Fig.~\ref{corr}.

Looking at interactions decreasing the bar strength, we did not find any particular parameter that seemed to be
responsible for this bar decrease. The orbital angles of the perturber in these cases do not seem to show a
preference for any value, but our sample may be too small (18 galaxies) to look for a consistent trend of a given
parameter to cause the bar decrease, especially if we restrict ourselves to flybys (12 cases). In the case of prograde
perturbers, it might be due to the position angle of the bar with respect to the perturber's position at the
pericentre. However, this is very difficult to probe in Illustris given the time resolution.

The fact that bars tend to decrease in strength over time, leading to a significant number of bar
disappearances, is interesting and puzzling, since we usually expect bars to grow stronger over time, as shown in
simulations of isolated galaxies (e.g. \citealt{2013MNRAS.429.1949A}). However, in a cosmological context this scenario
may not hold and it would be interesting to conduct this study in other cosmological simulations, such as the new
IllustrisTNG (\citealt{2018MNRAS.475..648P}).
This is however in contradiction with the results of EAGLE
(\citealt{2017MNRAS.469.1054A}) where the bars tend to grow stronger over time. This may indicate again a problem in
the bar formation and evolution in Illustris.

\citet{2010ApJ...714L.260N} found a bimodal distribution of the bar fraction as a function of stellar mass, with the
fraction increasing both at high and at low masses, with a minimum at intermediate masses. However, if we
restricted this bimodal distribution to the mass range we consider in the left panel of Fig.~\ref{frac_m} ($M_*>8.3
\times 10^{10}M_{\odot}$), we would only see the increasing trend towards higher masses. Therefore our results are
consistent with this bimodal distribution.

\section{Summary and conclusions}
\label{conclusion}

We studied barred galaxies in the Illustris simulation, with the emphasis on bars formed from flyby interactions. We
constructed a sample of disc galaxies in Illustris with masses higher than $M_*>8.3 \times
10^{10}M_{\odot}$ (100 000 stellar particles) and found that 21\% of them are barred. This fraction is lower than in observations, possibly
because of a too high softening length in Illustris, preventing the spontaneous formation of bars, or because of the issues related to the baryonic physics. We derived the bar
strengths, lengths and pattern speeds of these barred galaxies and found our values to be in the lower end of observations for the strength and pattern speed. The bar fraction in
Illustris disc galaxies increases with the total stellar mass and decreases with the amount of gas in the disc,
consistently with observations. Below $8.3 \times 10^{10}M_{\odot}$ we find almost no bars
at redshift $z=0$ in disc galaxies, which could point out to a resolution issue for low and intermediate mass galaxies,
in agreement with recent work on Illustris. We find the bar fraction to increase with redshift, which is in contradiction with observations, and the bars at higher
redshifts (up to $z=2$) to be often perturbed and asymmetric, so that the bar length and bar strength also tend to
increase with redshift.

We investigated the time evolution of bars to understand why we have less bars at low
redshift, and found that the main mechanism responsible for bar disappearance seems to be secular shrinking of the
bars, without external perturbations involved. Looking at the origin of the bars at redshift $z=0$, we found that very
few were formed in secular evolution, while most of them were created in mergers or flyby events. Illustris
thus forms bars at high redshift mostly from interactions, but those bars then slowly disappear over time in the
secular evolution of the galaxy, which leads to less numerous and weaker bars at low redshift.

We then focused on the effect of flybys on the bar formation and evolution. We selected a sample of 121 galaxies
undergoing a flyby interaction, which either creates a bar or affects the strength of a pre-existing bar. We found
that the corresponding change in the bar strength happens right after the pericentre passage. These tidally affected
bars tend to be stronger than average. We interpolated the orbit of the
perturber around the primary galaxy close to the pericentre, to obtain better precision in the values of the distance
and velocity. We then investigated different parameters of these interactions, in particular their strength and their
orbital angle, i.e. the angle between the perturber's orbit and the disc plane of the primary. We characterized the
strength of the interaction with the Elmegreen strength parameter. We found that the preferred scenario to create a bar
or to enhance it is with a perturber on a prograde orbit and interacting strongly (e.g. with a small pericentre
distance or a big mass for the perturber). Moreover, in interactions leading to the creation of a bar, the strength of
the created bar is proportional to the strength of the interaction.




\section*{Aknowledgements}
We thank Ivana Ebrova, Grzegorz Gajda, Klaudia Kowalczyk, Jean-Baptiste Salomon and Marcin Semczuk for
discussions and comments. We are grateful to the anonymous referee for useful suggestions that helped to
improve the paper. This work was supported by the Polish National Science Centre under grant 2013/10/A/ST9/00023.

\bibliographystyle{mnras}
\bibliography{biblio}


\appendix

\section{Uncertainty in mass estimates}
\label{app_mass}

\begin{figure}
  \includegraphics[scale=0.25]{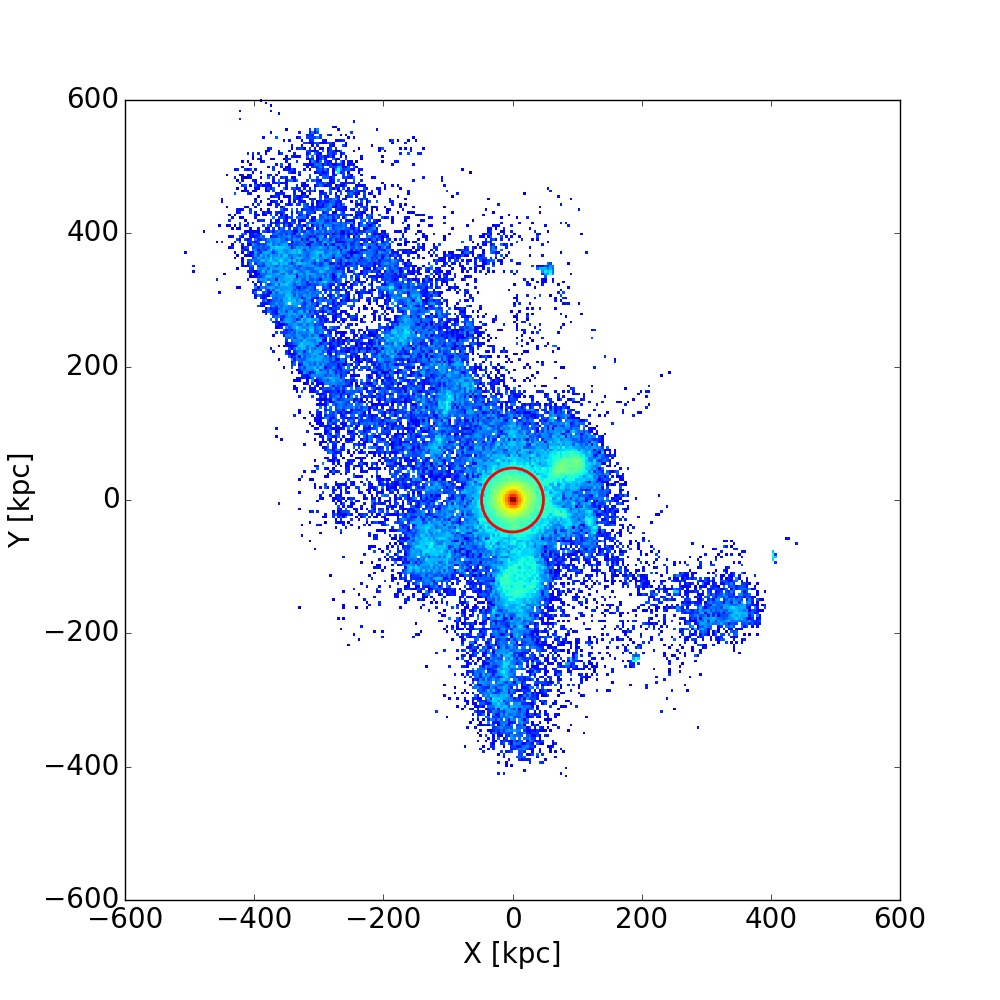}
  \caption{Example of a disc galaxy with wrongly attributed stellar particles. The image shows stellar density
distribution in the face-on view. The red circle coresponds to 10 times the scalelength of the exponential fit of the
stellar surface density profile.}
\label{stpart}
\end{figure}

In this Appendix we demonstrate how important it is to recompute galaxy masses from particle distributions, instead of
taking the masses provided by Illustris. We show an example of the stellar distribution of a disc galaxy seen face-on
in Fig.~\ref{stpart}, where we take all the stars Illustris defined as being part of this galaxy (called
\textit{subhalo}). One can see that although the galactic disc seems to extend only up to $\sim$ 50 kpc, there are many
stars that were attributed to this galaxy that are clearly outside the galaxy, up to 650 kpc away. Some of those stars
are stars from other nearby galaxies wrongly attributed to our galaxy, as one can see from the smaller stellar
concentrations within 200 kpc of the main galaxy.

Those galaxies are in interaction, which confuses the Illustris algorithm to attribute the stars. There are
also stars further away that are probably remnant of mergers or interactions, and are still attributed to our main
galaxy. All those stars do not seem relevant to take into account for our study, as they are not part of the galactic
disc, both dynamically and from their positions, and this is why we decided to remove them. As described in
section~\ref{barstr}, we thus fit the stellar surface density profile with an exponential, and keep only the stars
within 10 times the corresponding scalelength. The latter corresponds to the red circle in Fig.~\ref{stpart}. In this
example, this gives us a total stellar mass of $2.05 \times 10^{11}M_{\odot}$, instead of $2.61 \times
10^{11}M_{\odot}$ if we take all the stars attributed to this galaxy by Illustris, which corresponds to a decrease of
about 22 \%.

Note that to do our exponential fit we do not use the whole stellar distribution in the surface density
profile, since this would mean taking into account the stars that might be wrongly attributed. We take all the stars
within 5 times the half mass radius as provided by Illustris, use those stars to compute our own half-mass radius, and
then derive and fit the stellar surface density profile within 2 times this half-mass radius.

\section{Tests of the Tremaine-Weinberg method}
\label{app_TW}

\begin{figure}
  \includegraphics[scale=0.3]{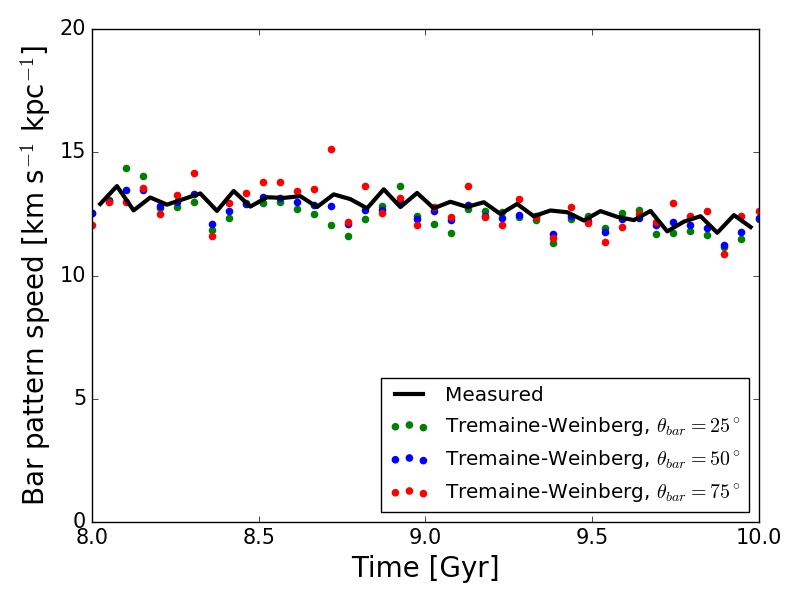}
  \caption{Bar pattern speed as a function of time for a higher resolution simulation of a barred galaxy. For the
black line the pattern speed is measured directly from the bar position angle over time, while the dots represent the
Tremaine-Weinberg measurements of the pattern speed, where the bar is rotated by different angles.}
\label{test_omp}
\end{figure}

In this Appendix we describe how we verified the reliability of the Tremaine-Weinberg method to derive the bar pattern
speed, using a higher resolution simulation. The simulation was taken from \citet{2016ApJ...826..227L} and involved
a Milky Way-like spiral galaxy built in isolation with a NFW dark halo and an exponential disc, with each component
having $10^6$ particles. The halo had a virial mass of $7.7 \times 10^{11}M_{\odot}$ and concentration of 27, while the
disc had a mass of $3.4 \times 10^{10}M_{\odot}$, a scalelength of 2.82 kpc and a thickness of 0.44 kpc. The simulation
was run for 10 Gyr and started to develop a bar after about 3 Gyr. We take the last 40 consecutive outputs of this
simulation, spaced by 0.05 Gyr, when the bar is already well developed, and measure the bar angle in each output. We
then derive the bar pattern speed by taking the angle difference between two snapshots. While this is
impossible to do in Illustris given the low time resolution, it can be done for this simulation given the short time
between snapshots, which is such that the bar rotates much less than 180 degrees during this period. We use this
pattern speed as a reference to which we compare the pattern speed derived with the Tremaine-Weinberg
method. We compute the latter using equation~(\ref{eqTW}), with the same parameters as in section \ref{patt}, and try
different values of the bar position angle. We then compare both methods in Fig.~\ref{test_omp}, by plotting the
pattern speed as a function of time. We see that the agreement  is very good, both methods giving results which
are similar within 10\%.

\end{document}